\documentclass[a4paper,11pt]{article}
\pdfoutput=1 
\usepackage{jheppub} 
\usepackage{bm}
\usepackage{slashed}
\usepackage[T1]{fontenc} 
\usepackage{graphicx}
\usepackage{subcaption}
\newcommand{\ri}{{\rm i}}

\title{\boldmath  Curvature-Assisted Dynamical Compactification in a Pre-Inflationary Higher-Dimensional Universe}

\author[a]{Yusuke Yamada}

\affiliation[a]{Cosmology, Gravity, and Astroparticle Physics Group, Center for Theoretical Physics of the Universe,
Institute for Basic Science (IBS), Daejeon, 34126, Korea}

\emailAdd{yamada@ibs.re.kr}

\abstract{
We investigate a pre-inflationary dynamical compactification scenario in which a higher-dimensional expanding universe evolves into an effectively four-dimensional one through curvature-assisted modulus trapping. To obtain a calculable semiclassical realization, we consider a simple five-dimensional $S^1$ compactification in a time-dependent open FRW background. Bulk quantum fields generate both the late-time Casimir contribution that stabilizes the radion and the early-time Kaluza--Klein thermal contributions relevant for the cosmological evolution. We show that negative curvature can sustain tracker-like radion evolution, allowing the radion to be trapped in a compactified vacuum before a subsequent four-dimensional inflationary phase dilutes the curvature remnant. While our analysis is performed in a 5D toy model, it illustrates a broader mechanism by which dynamical compactification can arise in a pre-inflationary higher-dimensional cosmology.}

\begin{document} 
\maketitle
\flushbottom
\section{Introduction}

Higher-dimensional spacetime is one of the most natural extensions of the standard model of particle physics suggested by ultraviolet completions of gravity such as string theory.
If extra dimensions exist, they must be hidden from present-day observations, and this makes compactification an important issue in connecting high-energy theory to low-energy physics.
It is true that there are alternative viewpoints, such as brane-world scenarios, in which the Standard Model degrees of freedom are localized on branes and the role of compactification may appear less central.
Nevertheless, in many attempts to derive realistic particle spectra from more fundamental constructions, the structure of the compact space still plays an essential role.
In this sense, it is not obvious that a purely brane-localized description of all observed physics is the unique or even the most natural answer, and compact extra dimensions may well remain an indispensable ingredient of realistic model building \cite{Ibanez:2012zz}.

Once one turns to cosmology, however, the issue of compactification becomes considerably more nontrivial.
By analogy with the observed three-dimensional universe, one natural expectation is that the early universe was characterized by very high energy scales, or equivalently by very short characteristic wavelengths, and hence by a small physical size.
If one applies the same logic to the extra dimensions, it is natural to consider the possibility that the compact directions were also dynamical and evolved from a small initial size.
Indeed, if the early universe contained sufficiently energetic matter, for example a hot thermal state, then the higher-dimensional spacetime as a whole would generically tend to expand.
The difficulty is that, in many cases, there is no efficient force that stops the expansion of the extra dimensions once it starts.

This problem is closely related to the Dine--Seiberg type behavior of moduli potentials:
in many compactifications the potential asymptotes to zero in the large-volume limit~\cite{Dine:1985he}.
Even when a stabilization mechanism exists, it is often achieved only through a local minimum generated by combining several effects, while the runaway behavior toward large volume remains.
On the other hand, the small-volume region is typically associated with a large effective energy density.
From the four-dimensional viewpoint, the cosmological expansion of the extra dimensions therefore corresponds to a modulus rolling down from a steep region of the potential toward a shallow local minimum.
In many situations the modulus acquires too much kinetic energy and overshoots the barrier instead of being trapped (see Fig.~\ref{fig:overshooting_problem}).
This is the overshooting problem, which has long been recognized as a serious obstacle to dynamical compactification \cite{Brustein:1992nk}.

One possible way to avoid overshooting is the tracker solution~\cite{Wetterich:1987fm,Ferreira:1997au,Copeland:1997et}.
If the dominant background component redshifts sufficiently slowly compared with the modulus kinetic and potential energy, the associated Hubble friction can force the modulus toward an attractor-like evolution in which its kinetic energy remains under control.
As a result, even a shallow local minimum may trap the field.
The general idea of tracker-assisted moduli stabilization is not new and has been discussed in various models \cite{Barreiro:1998aj,Huey:2000jx,Barreiro:2000pf,Brustein:2004jp,Kaloper:2004yj,Barreiro:2005ua,vandeBruck:2007jw,Barreiro:2007hb,Conlon:2008cj,Conlon:2022pnx,Alam:2022rtt,Apers:2024ffe,Brunelli:2025eif}.
Moreover, it has recently been emphasized that negative spatial curvature can serve as an especially efficient slowly redshifting component for this purpose \cite{Tonioni:2024huw}, which is remarkable in the sense that the curvature energy decays more slowly than dust or radiation energy often considered in the literature.

It is also known, in a different context, that open-universe curvature can ameliorate overshooting for the inflaton~\cite{Freivogel:2005vv,Dutta:2011fe} after Coleman--de Luccia tunneling \cite{Coleman:1980aw}.
In such scenarios, the negative curvature inside the bubble provides a large friction term that slows the inflaton before it reaches the inflationary plateau.
Thus, the use of curvature as a source of Hubble friction for a steeply rolling scalar field is itself not the novelty of the present work.

What we wish to study instead is a more specific cosmological question:
can a genuinely higher-dimensional, pre-inflationary expanding universe dynamically evolve into an effectively four-dimensional one through curvature-assisted modulus trapping, and can this happen in a way consistent with the subsequent inflationary history of our universe?
This question is distinct from the open-inflation overshooting problem because the scalar field of interest is not the inflaton evolving on a prescribed four-dimensional background, but the modulus controlling the size of the compact space.
The issue is therefore not merely to slow a rolling scalar field, but to achieve dynamical compactification into a metastable or stabilized four-dimensional vacuum in the presence of competing higher-dimensional sources.

This question is motivated by the fact that negative curvature is an efficient source of tracker behavior because it redshifts only as $a^{-2}$, while at the same time the present universe is observationally close to spatial flatness.
A natural way to reconcile these facts is to assume that curvature plays an important role before inflation, while a later four-dimensional inflationary phase dilutes the curvature remnant after the modulus is stabilized.
Since inflation itself is strongly supported observationally, such a sequence,
\[
\text{higher-dimensional expansion}
\;\to\;
\text{curvature-assisted tracker}
\;\to\;
\text{modulus stabilization}
\;\to\;
\text{4D inflation},
\]
(see also Fig.~\ref{fig:scenario_concept}) is both phenomenologically motivated and conceptually coherent.
We note that higher-dimensional cosmology has been investigated in the literature, particularly since the 1980s~\cite{Chodos:1979vk,Randjbar-Daemi:1983awk,Shafi:1983hj,Kolb:1983fm,Sahdev:1984gts,Shafi:1984ha,Okada:1984cv,Maeda:1984gq,Abbott:1984ba,Maeda:1985bq,Maeda:1986gh}.

There is also an important structural reason why curvature is useful in this problem.
In the Einstein-frame description of product-type compactifications, the spatial curvature of the non-compact slices appears primarily as a background contribution to the Friedmann equation, rather than as an additional model-dependent radion potential.
As a result, it can enhance Hubble friction and support tracker-like evolution without, at leading order, deforming the stabilizing potential itself.
This is qualitatively different from matter sources such as thermal excitations or excited Kaluza--Klein (KK) modes, whose contributions generally induce radion-dependent effective terms and may erase the local minimum before the modulus reaches it.
In this sense, negative curvature is not only efficient because it redshifts slowly, but also because it provides a comparatively robust source of friction that is less sensitive to the detailed structure of the matter sector.

One may nevertheless wonder whether a question of this kind should instead be addressed directly in a fully string-theoretic description.
Indeed, in sufficiently high-energy regimes, the relevant degrees of freedom may no longer be those of semiclassical gravity and bulk quantum fields.
Nevertheless, whenever there exists an intermediate regime in which the higher-dimensional background and its excitations admit a controlled field-theoretic description, the dynamical compactification problem should be analyzed within that regime itself.
Our aim in this paper is precisely to study such a controllable semiclassical regime.
In particular, as discussed in Appendix~\ref{appC}, we restrict attention to a parameter range in which the thermal, KK, and curvature scales remain below the appropriate higher-dimensional cutoff, so that the semiclassical treatment is self-consistent.
In this sense, our analysis is complementary to, rather than in competition with, more intrinsically stringy descriptions of the very early universe (see e.g.~\cite{Gasperini:2002bn,Brandenberger:2011et}).

The main purpose of this paper is to present an explicit realization of this question in a simple but calculable five-dimensional toy model.
We consider a five-dimensional spacetime compactified on $S^1$, in which bulk quantum fields play two roles simultaneously:
they generate the late-time Casimir contribution that stabilizes the radion, and they also provide the early-time thermal and KK energy contributions relevant for the cosmological evolution.
In this sense, our setup goes beyond a purely phenomenological fluid approximation.
Instead, it allows one to derive the effective source terms entering the semiclassical Einstein equations directly from quantum field theory in a time-dependent higher-dimensional background.

This calculable semiclassical framework is useful for two reasons.
First, it provides a concrete realization of the idea that a pre-inflationary higher-dimensional era can dynamically connect to an effectively four-dimensional universe.
Second, it makes it possible to track radion- and background-dependent effective terms that would be invisible in a simple fluid model.
In particular, the same quantum effects that are responsible for late-time stabilization can also modify the early-time dynamics in nontrivial ways.
Depending on the regime, such effects may assist the tracker behavior, but they may also obstruct it by distorting the effective potential or by making the stabilized minimum shallower or even unstable.
For example, sufficiently large curvature- or Hubble-induced contributions can increase the effective mass of bulk scalar fields and thereby change the vacuum contribution to the radion potential.
Although such effects do not spoil the successful tracking solutions in the parameter range studied in this paper, the calculable model teaches us that the consistency of tracker-assisted compactification is not automatic and must be checked dynamically within the same framework.

To state the novelty more precisely, the present work does not claim novelty for the general idea that curvature friction can slow a scalar field.
Rather, to the best of our knowledge, the new element is the following combination.
First, we formulate a pre-inflationary transdimensional scenario in which a genuinely higher-dimensional expanding universe dynamically evolves into an effectively four-dimensional one through curvature-assisted radion trapping, instead of assuming already stabilized extra dimensions as part of a four-dimensional effective description.
Second, we realize this idea in an explicit five-dimensional semiclassical model in which the same bulk quantum fields generate both the late-time Casimir stabilization and the early-time thermal/KK source terms.
Third, the model allows us to test dynamically whether the very effects responsible for stabilization remain compatible with the preceding cosmological evolution.
In this way, the paper addresses not only scalar overshooting, but the consistency of dynamical compactification itself.
This also suggests a broader lesson:
the strategy developed here may be extendable, at least in principle, to more general compactifications and higher-dimensional settings, where calculable quantum effects may again play a decisive role in determining whether dynamical compactification succeeds or fails.

This paper is organized as follows.
In Sec.~\ref{sec:curvature_assisted_compactification}, we review the curvature-assisted tracker idea and illustrate it first in a four-dimensional example.
In Sec.~\ref{sec:5Dmodel}, we introduce the five-dimensional model and derive the semiclassical equations governing the radion and the cosmological background.
We briefly discuss quantization of matter fields in the background spacetime and compute the relevant quantum expectation values for bulk fields in the time-dependent background, including the vacuum and thermal contributions.
In Sec.~\ref{sec:4Demergence}, we study the resulting dynamics and show how an effectively four-dimensional universe can emerge from an expanding five-dimensional phase.
We conclude in Sec.~\ref{sec:conclusion} with a discussion of implications, limitations, and possible extensions.
In Appendix~\ref{quantum field in openFRW}, we summarize the quantization of scalar and spinor fields in our time-dependent curved background.
The energy density of bulk quantum fields in a thermal state is computed in Appendix~\ref{appB}.
We also discuss the cutoff scale within our model in Appendix~\ref{appC}, which is relevant to the choice of model parameters in our 5D model.
In Appendix~\ref{appD}, we comment on the behavior of scalar fields' effective frequency that appears in our quantization procedure.

Throughout this work, we take the convention $\hbar=c=k_B=1$.

\section{Curvature-assisted compactification before inflation}
\label{sec:curvature_assisted_compactification}

It is not known how a higher-dimensional spacetime motivated by string theory
evolved into the four-dimensional universe we observe today. A natural
extrapolation of the hot Big-Bang picture is that the earliest universe was hot
and genuinely higher-dimensional, with both the non-compact dimensions and the
extra dimensions undergoing cosmological evolution. In such a picture, the extra
dimensions must eventually stop expanding and become stabilized; otherwise the
late universe would not admit an ordinary four-dimensional effective
description. 

\begin{figure}[t]
  \centering
   \includegraphics[width=0.85\linewidth]{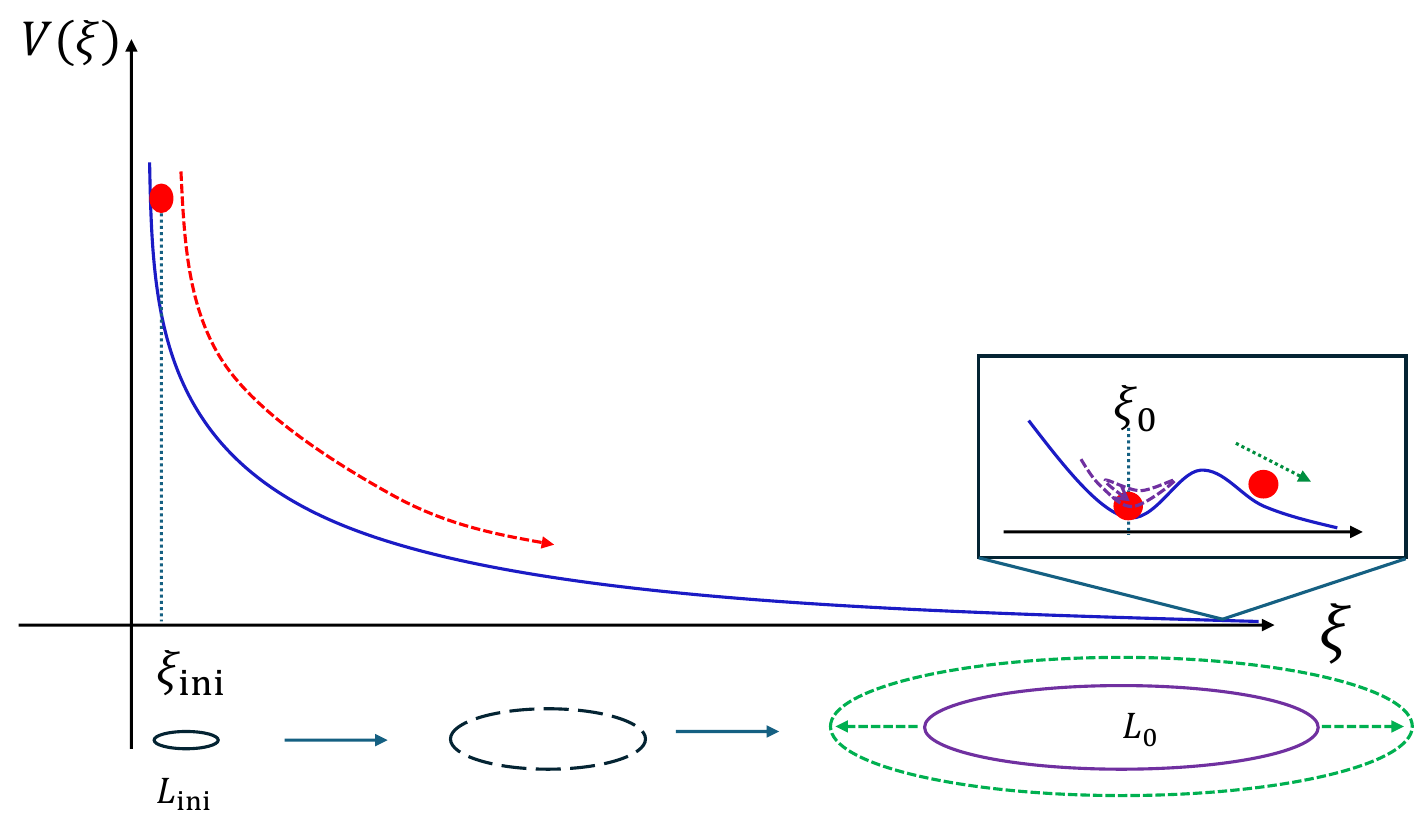}
  \caption{ Illustration of the overshooting problem. The expansion of the extra dimension is equivalent to the dynamical evolution of a volume modulus field $\xi$, whose VEV determines the extra dimension volume $L$. A local minimum exists, and if a modulus is trapped by the barrier, the extra dimension volume can be stabilized. 
  However, a modulus rolling from a steep region can acquire too much kinetic energy
  and pass over the shallow barrier. The compact dimension then runs away rather
  than settling at the stabilized size as sketched in the green dashed line.  }
  \label{fig:overshooting_problem}
\end{figure}

Even if a compactification potential has a local minimum, this does not by
itself guarantee successful compactification. Suppose that the compactification
modulus, or radion, is denoted by $\xi$, and that the desired compactified
universe corresponds to a local minimum at $\xi=\xi_0$. If the compact space expands, or equivalently if the modulus starts to roll down far from the minimum, it can acquire kinetic energy larger than the height of the barrier
separating the minimum from the decompactified region. The modulus then passes
over the minimum and runs away. This is the overshooting problem, shown schematically in
Fig.~\ref{fig:overshooting_problem}. Thus, in a genuinely higher-dimensional
cosmology, it is not enough to prepare a static potential that has a stabilized
vacuum. One also needs a dynamical mechanism that reduces the modulus kinetic
energy before the field reaches the minimum.

One possible way to avoid the overshooting problem is through a tracker solution~\cite{Wetterich:1987fm,Ferreira:1997au,Copeland:1997et}, which has been applied to moduli stabilization in \cite{Barreiro:1998aj,Huey:2000jx,Barreiro:2000pf,Brustein:2004jp,Kaloper:2004yj,Barreiro:2005ua,vandeBruck:2007jw,Barreiro:2007hb,Conlon:2008cj,Conlon:2022pnx,Alam:2022rtt,Apers:2024ffe,Tonioni:2024huw,Brunelli:2025eif}. To see this, consider a canonically normalized modulus $\phi$ rolling on a steep
exponential potential,
\begin{equation}
  V(\phi)\simeq V_0 e^{-\lambda\phi/M_{\rm Pl}} .
\end{equation}
In the presence of a background fluid with equation-of-state $p=(\gamma-1)\rho$, the tracker solution gives
\begin{equation}
  \Omega_\phi \simeq \frac{3\gamma}{\lambda^2},
  \qquad
  \frac{\dot\phi^2/2}{V}
  \simeq
  \frac{\gamma}{2-\gamma}.
\end{equation}
The existence of the above tracker solution requires
\begin{align}
    \lambda^2>3\gamma.\label{trackercond}
\end{align}
The kinetic energy is therefore kept comparable to the potential energy, rather
than becoming arbitrarily large. If the field reaches the vicinity of the local
minimum while this tracker-like behavior is active, the modulus can be trapped
instead of overshooting the barrier.

Among the simple background components commonly considered in cosmology, negative curvature is particularly efficient in realizing the tracker solution, since the curvature contribution redshifts only as $a^{-2}$, much more slowly than matter $\propto a^{-3}$ or radiation $\propto a^{-4}$. In~\cite{Tonioni:2024huw}, the author pointed out that the curvature energy can universally drive moduli dynamics in which all moduli approach constant values and the universe approaches the Milne universe or equivalently the Minkowski spacetime. In the present work, we adopt this idea in a different setting, namely an open FRW cosmology in which the curvature-dominated stage is followed by radion stabilization and a subsequent four-dimensional inflationary phase. In the four-dimensional Einstein-frame description, the curvature
contribution is
\begin{equation}
  \rho_K \equiv -\frac{3KM_{\rm pl}^2}{a^2}, \qquad K<0,
\end{equation}
or equivalently it behaves as an effective fluid with
 $ \gamma_K =\frac{2}{3}$, which redshifts more slowly than radiation $\rho_r\propto a^{-4}$ and matter $\rho_m\propto a^{-3}$. Because it is diluted slowly, curvature can remain important for a long period
and enhance the Hubble friction acting on the modulus. This makes curvature
particularly efficient in realizing a tracker-like evolution. 

Another important advantage of curvature is more structural. In the Einstein-frame description of product-type compactifications, the spatial curvature of the non-compact slices appears primarily as a background contribution to the Friedmann equation, rather than as an additional model-dependent radion potential. As a result, it can enhance Hubble friction and support tracker-like evolution without, at leading order, deforming the stabilizing potential itself. This is qualitatively different from matter sources such as thermal excitations or excited KK modes, whose contributions generally induce radion-dependent effective terms and may erase the local minimum before the modulus reaches it. In this sense, negative curvature is not only efficient because it redshifts slowly, but also because it provides a comparatively robust source of friction that is less sensitive to the detailed structure of the matter sector.

\begin{figure}[t]
  \centering
 \includegraphics[width=1.0\linewidth]{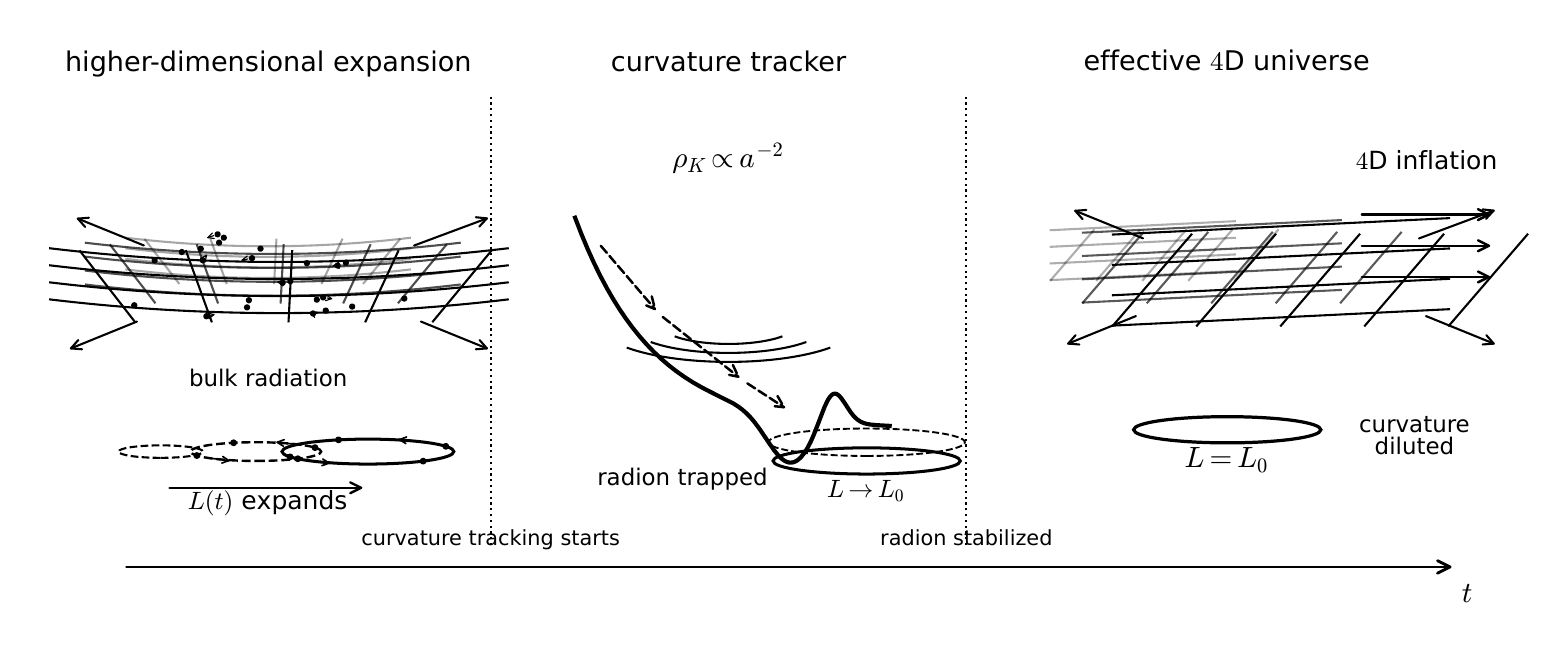}
  \caption{
  Illustration of our scenario. The universe starts from a
  higher-dimensional expanding phase with dynamical compact dimensions. Bulk
  radiation can carry momentum both along the non-compact directions and around
  the compact space. In the early phase, the compact space volume $L$ increases, which means that the modulus rolls down a steep potential. Nevertheless, the negative curvature component, redshifting as
  $\rho_K\propto a^{-2}$, can drive a tracker-like modulus evolution. After the
  modulus is stabilized at $L=L_0$, a subsequent four-dimensional inflationary
  phase dilutes the curvature contribution.
  }
  \label{fig:scenario_concept}
\end{figure}

The scenario we have in mind is summarized in Fig.~\ref{fig:scenario_concept}.
The universe begins in a higher-dimensional expanding phase. Both the
non-compact scale factor and the size of the compact space are time dependent.
Bulk radiation or excited KK modes may also be present. In this
stage, the radion evolves in a cosmological background that is not yet an
ordinary four-dimensional effective universe.

The role of negative curvature is to provide a slowly redshifting background
component that can dynamically stop the expansion of the compact dimensions. If
the modulus is trapped at the local minimum, the universe enters an effective
four-dimensional phase. However, the curvature-dominated stage cannot persist
until late times, since observations require the present universe to be close to
spatially flat. We therefore assume that the modulus trapping is followed by a
four-dimensional inflationary phase, which dilutes the curvature contribution
and connects the compactified universe to the standard hot Big-Bang cosmology.

We also note that our scenario differs from models in which the compact dimensions grow after inflation \cite{Arkani-Hamed:1999fet,Conlon:2008cj,Alam:2022rtt,Apers:2024ffe,Conlon:2022pnx,Brunelli:2025eif}, or from scenarios in which the compact space inflates together with the three large spatial dimensions \cite{Anchordoqui:2023etp,Antoniadis:2023sya,Anchordoqui:2024amx,Hirose:2025pzm}. The underlying issue is similar in that late-time volume stabilization must still overcome overshooting, but in the present setup open-universe curvature provides a particularly efficient way to realize tracker-like evolution while remaining observationally viable once a subsequent four-dimensional inflationary phase dilutes the curvature remnant.

From a phenomenological perspective, within our scenario, the early higher-dimensional phase may leave KK particles and open universe curvature, which may affect the behavior of the cosmological curvature perturbation, provided that the observable scales leave the horizon when these remnant effects are not completely erased by the accelerated expansion. We do not address the observational signature in this work, but leave it as an interesting future direction.

\subsection{Warm-up example: KKLT modulus}

Before studying an explicit higher-dimensional model, it is useful to illustrate
the same mechanism in a familiar four-dimensional potential. We consider the
Kachru-Kallosh-Linde-Trivedi (KKLT)-type model~\cite{Kachru:2003aw} with a volume modulus
\begin{equation}
  T=\sigma+i\tau ,
\end{equation}
with K\"ahler potential and superpotential
\begin{equation}
  K=-3M_{\rm pl}^2\log(T+\bar T),
  \qquad
  W=W_0+A e^{-\alpha T},
\end{equation}
and the potential in supergravity is given by
\begin{align}
    V=e^{\frac{K}{M_{\rm pl}^2}}(K^{T\bar{T}}|D_TW|^2-3|W|^2/M_{\rm pl}^2)
\end{align}
where $K^{T\bar T}=(\partial_T\partial_{\bar T}K)^{-1}$, $D_TW=\partial_T W+M_{\rm pl}^{-2}\partial_TK W$. In the KKLT model, one adds the uplifting term originating from an anti-D3 brane
\begin{equation}
  V_{\rm up}=\frac{C}{\sigma^3}
\end{equation}
with a constant $C$. For this illustrative example, we fix the axion $\tau$ at an extremum, $\alpha\tau=0$,
and focus only on the real part $\sigma$. The canonically normalized field is
\begin{equation}
  \phi=\sqrt{\frac{3}{2}}M_{\rm pl}\log\sigma .
\end{equation}
The modulus $\sigma$ is related to the six-dimensional compact-space volume ${\cal V}_6$ as $\sigma\sim ({\cal V}_6)^{\frac23}$.

The dynamics in the presence of negative curvature is described by
\begin{align}
  &\ddot\phi+3H\dot\phi+\frac{\partial V_{\rm KKLT}}{\partial\phi} = 0,\\
  &3M_{\rm pl}^2H^2 = \frac{1}{2}\dot\phi^2+V_{\rm KKLT}(\phi)+\rho_K,
\end{align}
where $\rho_K\propto a^{-2}$ denotes the energy density of the negative curvature. The curvature term does not directly modify the shape of the KKLT potential, but increases the Hubble friction through the Friedmann equation.

\begin{figure}[t]
  \centering
  \begin{subfigure}{0.48\linewidth}
    \centering
    \includegraphics[width=\linewidth]{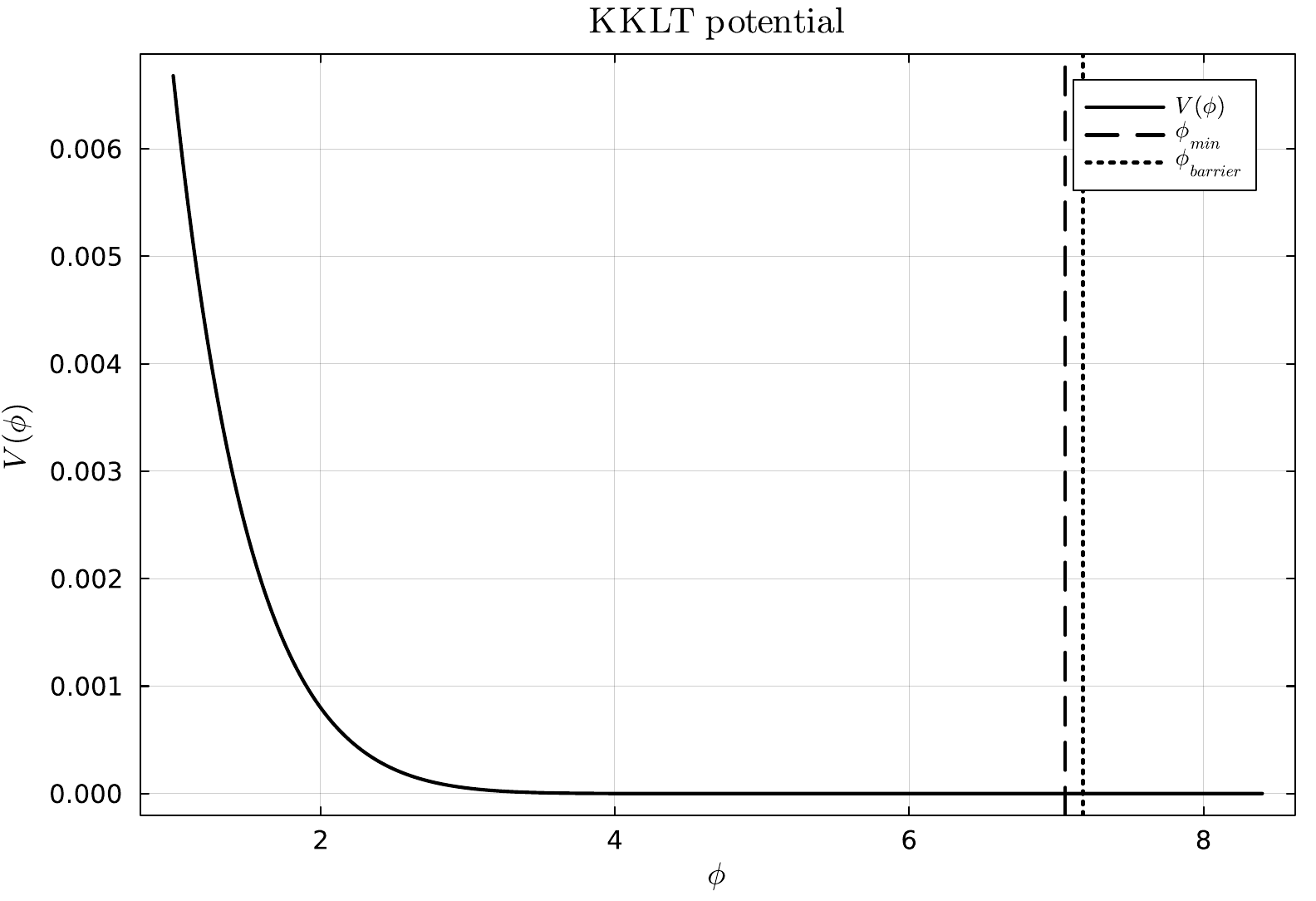}
    \caption{Linear scale.}
  \end{subfigure}
  \hfill
  \begin{subfigure}{0.48\linewidth}
    \centering
     \includegraphics[width=\linewidth]{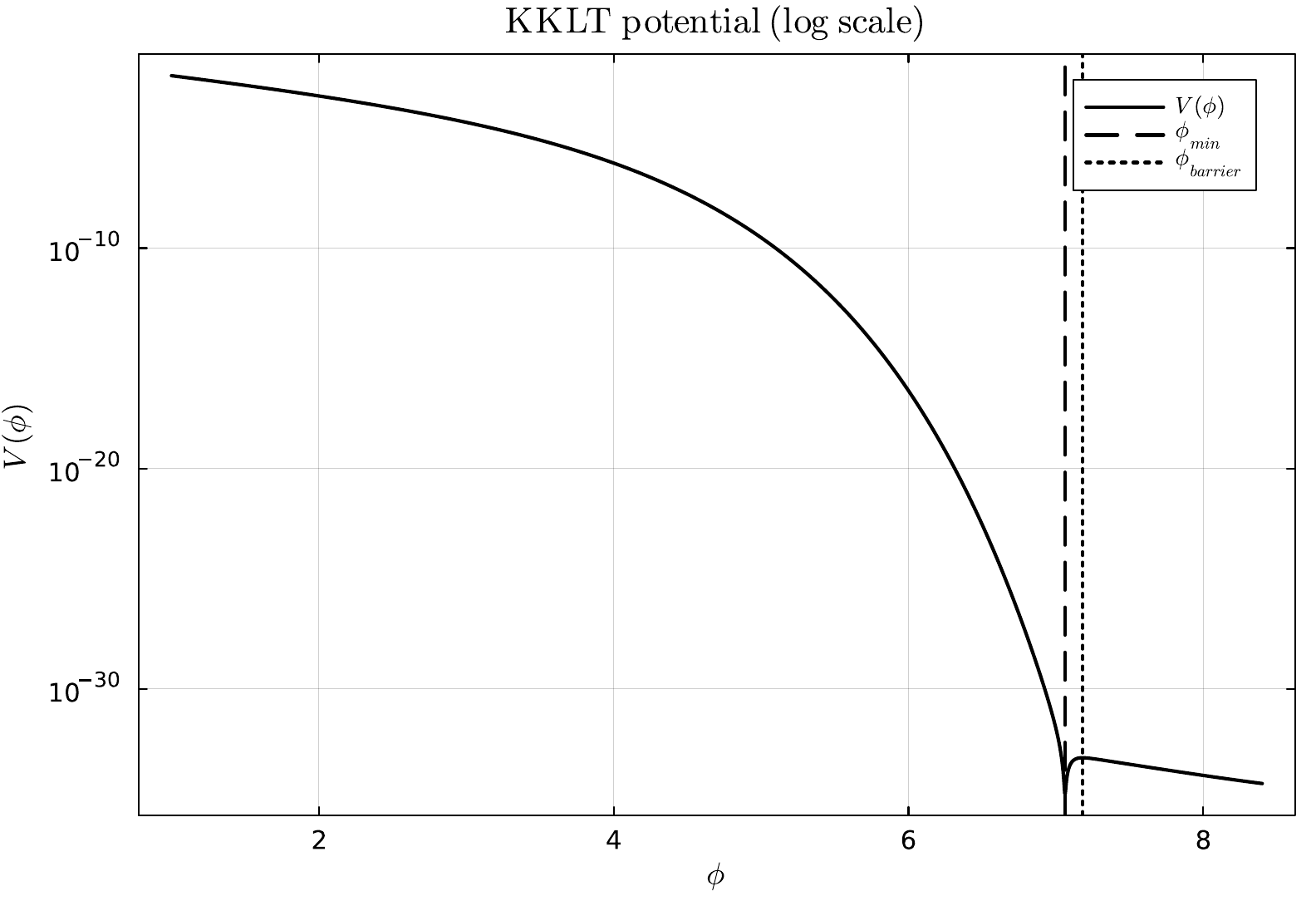}
    \caption{Log scale.}
  \end{subfigure}
  \caption{
  KKLT potential with a fixed imaginary part, which shows the local
  structure near the stabilized region, while the logarithmic plot emphasizes
  the hierarchy between the steep region and the shallow barrier.
  }
  \label{fig:kklt_potential}
\end{figure}

The potential in Fig.~\ref{fig:kklt_potential} explicitly shows how nontrivial the tracker solution is, since the modulus rolls down from a hierarchically larger value of the potential to an extremely shallow local minimum. Naively, the field excursion from $\phi\sim M_{\rm pl}$ would never be stopped at the local minimum around $\phi\sim 7M_{\rm pl}$. Note also that $\phi\to \infty$ is the large volume limit in terms of the compact spaces, and the potential asymptotes to zero within this setup. This is related to the so-called Dine-Seiberg problem~\cite{Dine:1985he}.

\begin{figure}[t]
  \centering
  \includegraphics[width=0.7\linewidth]{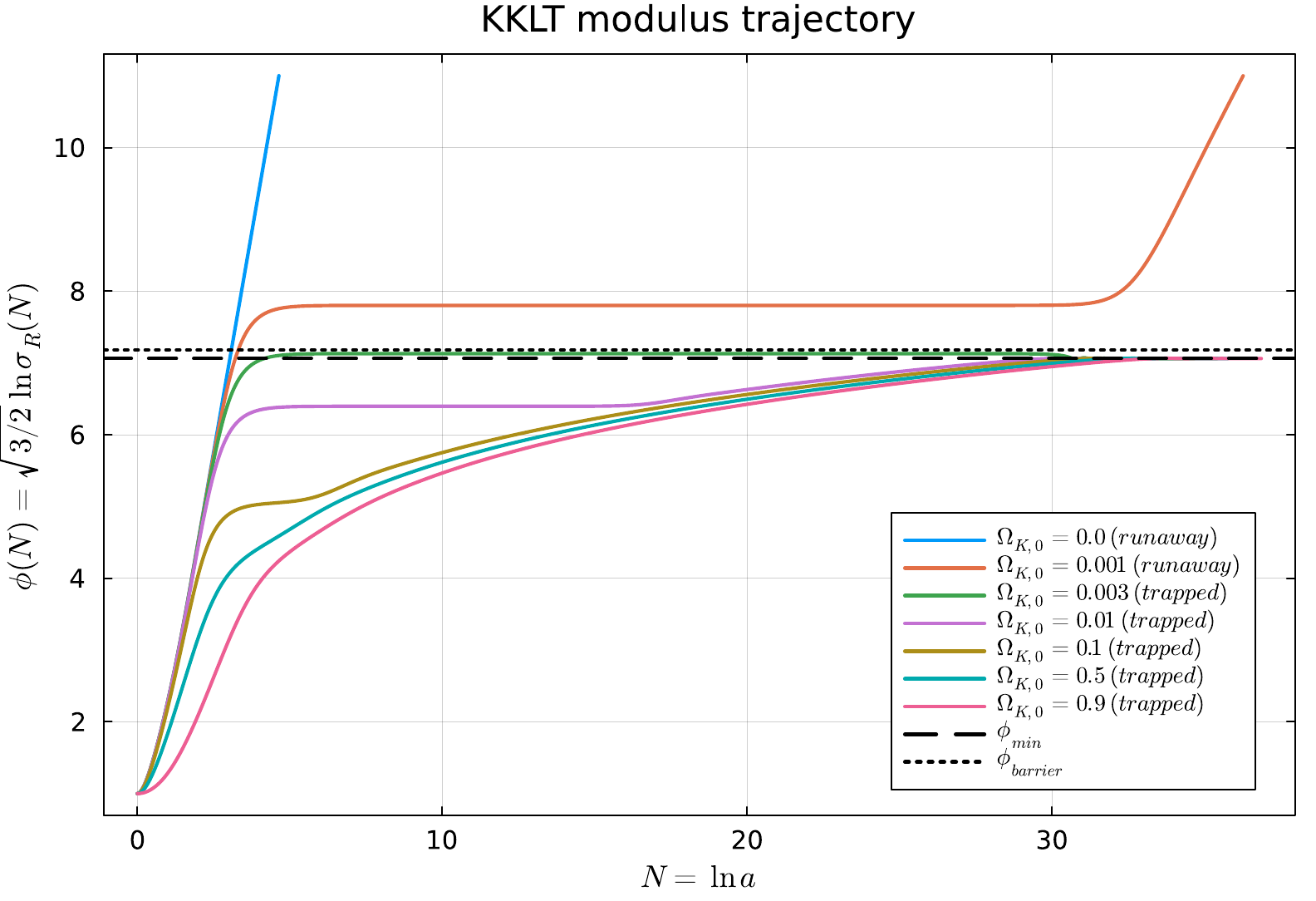}
  \caption{
  Evolution of the canonically normalized KKLT modulus
  $\phi(N)=\sqrt{3/2}M_{\rm pl}\ln\sigma_R(N)$ for different initial curvature fractions.
  For small curvature fraction, the modulus overshoots the barrier and runs
  away. For sufficiently large curvature fraction, curvature-enhanced Hubble
  friction slows the field down and the modulus is trapped near
  $\phi_{\rm min}$.
  }
  \label{fig:kklt_modulus_vs_N}
\end{figure}

Figure~\ref{fig:kklt_modulus_vs_N} shows a simple numerical realization of the curvature-assisted trapping mechanism. For small initial curvature fraction, the field crosses the barrier and runs away. Increasing the initial curvature fraction enhances Hubble friction and allows the field to settle near the local minimum. This example is not intended as a complete early-universe model, but it makes the dynamical role of curvature transparent. In earlier works~\cite{Huey:2000jx,Brustein:2004jp,Barreiro:2005ua,Barreiro:2007hb,vandeBruck:2007jw}, the background fluid is considered as either radiation or matter, but the curvature decays slower than those components, and therefore, the curvature can be a more efficient ingredient for realizing tracker-like evolution.

For realistic model building, this is not the end of the story since the resulting curvature-dominated universe is observationally inconsistent. However, as we discuss in the 5D example below, if inflation occurs after the dynamical radion stabilization, the curvature energy can be diluted.

The KKLT example is only a four-dimensional illustration, but it makes the dynamical role of curvature transparent and serves as a useful warm-up for the higher-dimensional model studied below.

\section{A Semiclassical 5D Model of Dynamical Compactification}\label{sec:5Dmodel}
In this section, we construct a simple five-dimensional semiclassical model that realizes the pre-inflationary dynamical compactification scenario discussed above. The setup is still a toy model, but it is sufficiently explicit to allow a first-principles treatment of both the vacuum effects responsible for radion stabilization and the thermal/KK contributions relevant for the early cosmological evolution. Our goal is to formulate the coupled dynamics of the background geometry, the radion, and bulk quantum matter in a single calculable framework.

Although the model is only a five-dimensional toy example, it captures the essential ingredients of the mechanism we wish to illustrate: a higher-dimensional expanding phase, a radion potential generated by bulk quantum effects, and early-time matter sources that can either assist or obstruct successful trapping. Negative curvature can
provide a slowly redshifting component that drives tracker-like radion
evolution. Once the radion is trapped, a later four-dimensional inflationary
phase can dilute curvature and lead to ordinary four-dimensional cosmology. Furthermore, both the thermal contributions relevant for the early universe and the quantum effects responsible for radion stabilization can be computed from first principles in a non-trivial background, which makes the scenario quantitatively testable within the same framework.

We consider the following toy model in 5D spacetime compactified on $S^1$:
\begin{align}
    S=&\int d^4x \int_0^{2\pi R} dy \sqrt{-g}\Biggl[\frac{M_5^3}{2}{\cal R}_5-\Lambda_5-\frac12\partial_M\Phi\partial^M\Phi-V(\Phi)\nonumber\\
    &-\frac12\sum_{i=1}^{N_B}(\partial_M\phi_i\partial^M\phi_i+m_i^2\phi_i^2)-\sum_{I=1}^{N_F}\left(\bar{\Psi}_I\Gamma^ae_a^MD_M\Psi_I-m_I\bar{\Psi}_I\Psi_I\right)\Biggr],
\end{align}
where $M_5$ is the 5D Planck scale, ${\cal R}_5$ is the 5D Ricci scalar, $\Lambda^5$ is the cosmological constant, $\Phi$ and $\phi_i$ ($i=1,2,\cdots)$ are real scalar fields, and $\Psi_I$ ($I=1,2,\cdots$) are Dirac spinors. 
We consider the background spacetime to be a 5D version of open FRW metric
\begin{align}
    ds^2=g_{MN}dx^Mdx^N=\frac{1}{b(t)}\left(-dt^2+a^2(t)\left(\frac{dr^2}{1+r^2/r_K^2}+r^2(d\theta^2+\sin^2\theta d\varphi^2)\right)\right)+b^2(t)dy^2,\label{5D metric}
\end{align}
with the periodic boundary condition $y\sim y+2\pi R$ where $R$ is a reference 5D radius. The parameter $r_K$ denotes the spatial curvature radius. The above metric keeps the 4D Planck scale to be constant $M_{\rm pl}^2\equiv 2\pi R M_5^3$, while the physical 5D radius is given by
\begin{align}
    L(t)=\int_0^{2\pi R} dy \sqrt{G_{yy}}=2\pi R b(t).
\end{align}
The real scalar $\Phi$ will be identified as the inflaton field, which drives inflation effectively after radion is stabilized at the local minimum. The matter fields $\phi_i$ and $\Psi_I$ play two roles simultaneously. First, they generate the Casimir contribution that stabilizes the radion at late times together with the cosmological constant. Second, when excited in the early universe, they provide radiation-like or matter-like energy contributions that affect the pre-inflationary dynamics. 

We focus on the dynamics of the radion field $b(t)$. Except the 5D cosmological constant, the radion has no potential at the tree level. However, bulk matter fields yield one-loop corrections to the vacuum energy, which is often called Casimir energy. We will explicitly compute such a contribution, but here we express it schematically as $V_{\rm loop}$. Taking into account the Casimir energy term, the radion effective action
\begin{align}
    S_{\rm rad}=&\int d^4xa^3\left[\frac34M_{\rm pl}^2\left(\frac{\dot b}{b}\right)^2-\frac{2\pi R\Lambda_5}{b}-V_{\rm loop}(b)\right]\nonumber\\
    =&\int d^4xa^3\left[\frac12{\dot\xi}^2-\Lambda_{4D}e^{-\sqrt{\frac23} \frac{\xi}{M_{\rm pl}}}-V_{\rm loop}(b(\xi))\right]
\end{align}
where we have canonically normalized the radion as $\xi=\sqrt{\frac32}M_{\rm pl}\log b$ and $\Lambda_{4D}=2\pi R\Lambda_5$. The Casimir energy contribution from massless fields is proportional to $b^{-6}=\exp\left(-2\sqrt6\xi/M_{\rm pl}\right)$, which dominates the potential as $b\to 0$. We consider a model where the Casimir energy and the cosmological constant term stabilize the radion at some field value in the present universe. However, in the early phase, there can be extra contributions. For instance, if the bulk matter fields are in an excited state, such as a thermal equilibrium state, additional (time-dependent) effective potential terms appear, which would vanish in the present universe.

In the following, we consider the early 5D Big Bang phase, which is supposed to be the pre-inflationary era. It should be emphasized that we do not discuss the origin of the differences of 3D spaces and $S^1$ direction, which is already assumed in the metric ansatz~\eqref{5D metric}. Rather, we ask how effectively 4D cosmology may emerge from 5D Big-Bang cosmology. This is yet a nontrivial question, in general due to the overshooting problem: Let $b_0$ denote the radion value at the local minimum. If we allow the compact space and 3D space to expand in the initial phase, the initial radion value $\xi_{\rm ini}$ is much smaller than $\xi_0$. From a 4D perspective, we may think of 5D Big-Bang dynamics as 4D spacetime with a radion scalar rolling down on the potential from $\xi_{\rm ini}$. The question is whether $\xi$ can be trapped at $\xi_0$ at some moment without overshooting to $\xi\to \infty$. 

\subsection{Equations of Motion}
In this section, we derive the equations of motion of the system. We treat gravitational fields and the inflaton field $\Phi$ as classical fields while treating matter fields as quantum fields. We note that the quantum fluctuation of $\Phi$ can be included, but strictly speaking, we need to expand $\Phi$ as the classical background part and fluctuation having background dependent masses, which complicates the discussion. Therefore, we do not include the inflaton fluctuation explicitly, but one may identify one of the quantum scalar fields as the inflaton (but the mass is approximated by a constant). We then evaluate the energy-momentum operator in the vacuum or (initial) thermal state, which gives the energy of the fluid and Casimir energy simultaneously. 

The Einstein equation describes the time evolution of the 3D universe as well as the radion. The $00$-component of the Einstein tensor reads 
\begin{align}
    G_{00}=3H^2-\frac12\frac{\dot\xi^2}{M_{\rm pl}^2}+\frac{3}{a^2r_K^2},
\end{align}
and the useful linear combination giving the radion equation of motion is
\begin{align}
    2G^5{}_5-\sum_{i=1}^3G^i{}_i-G^0{}_0=3\sqrt{\frac23}\left(\frac{\ddot\xi}{M_{\rm pl}}+3H\frac{\dot\xi}{M_{\rm pl}}\right)b,
\end{align}
and accordingly, we evaluate the expectation values of $\hat{T}_{00}$ and $\hat{T}^5{}_5-\sum_{i=1}^3\hat{T}^i{}_i-\hat{T}^0{}_0$ of matter fields. These operators are explicitly given by
\begin{align}
    \hat{T}_{00}=&\sum_{i=1}^{N_B}\left[\frac12\dot{\hat\phi}_i^2+\frac{1}{2}({\bm \nabla}\hat{\phi}_i)^2+\frac{1}{2b}\partial_y\hat\phi\partial^y\hat{\phi}_i+\frac{m^2_i}{2b}\hat\phi^2_i\right]-\sum_{I=1}^{N_F}\frac12\hat{\bar{\Psi}}_I\Gamma^ae_{a0}\overleftrightarrow{D}_0\hat{\Psi}_I+\frac{\Lambda_5}{b}\hat{\bm 1}\nonumber\\
    =&\hat\rho_{\phi}+\hat\rho_{\Psi}+\frac{\Lambda_5}{b}\hat{\bm 1},
\end{align}
and
\begin{align}
    2\hat{T}^5{}_5-\sum_{i=1}^3\hat{T}^i{}_i-\hat{T}^0{}_0=&\sum_{i=1}^{N_B}\left[3\partial_y\hat{\phi}_i\partial^y\hat\phi_i+m_i^2\hat{\phi}_i^2\right]+\sum_{I=1}^{N_F}\left[-\frac32\hat{\bar{\Psi}}_I\Gamma^a e_a^5 \overleftrightarrow{D}_5\hat{\Psi}_I+m_I\hat{\bar\Psi}_I\hat{\Psi}_I\right]+2\Lambda_5\hat{\bm1}\nonumber\\
    =&3\hat{J}_\phi+3\hat{J}_\Psi+2\Lambda_5\hat{\bm1},
\end{align}
where $A\overleftrightarrow{D}_MB=A(D_MB)-(D_MA)B$, we have used free Heisenberg equation of $\Psi$ in deriving both expressions, and $\hat{T}^M{}_N$ represents a tensor with mixed upper and lower indices. Accordingly, the semiclassical Einstein equation is given by
\begin{align}
    &3H^2M_{\rm pl}^2=\frac12\dot\xi^2+\frac{\Lambda_{\rm 4D}}{b}+\frac{3M_{\rm pl}^2}{a^2r_K^2}+2\pi R\left(\langle\hat\rho_\phi\rangle+\langle\hat{\rho}_\Psi\rangle\right),\label{EOM1}\\
    &\ddot\xi+3H\dot\xi-\sqrt{\frac23}\frac{\Lambda_{\rm 4D}}{M_{\rm pl}b}=\frac{2\pi R}{M_{\rm pl}b}\sqrt{\frac32}\left(\langle J_\phi\rangle+\langle J_\Psi\rangle\right),\label{EOM2}
\end{align}
where the expectation value $\langle\cdot\rangle$ is evaluated in a quantum state, which we will specify later. 

Thus, we have a set of equations that can be solved once we specify the initial conditions including the initial state in which the operators are evaluated. Note that the operators are in the Heisenberg picture, and they obey the Heisenberg equation or equivalently the mode equations, which we also need to solve. In the next subsection, we discuss how to evaluate the expectations of quadratic operators in the Einstein equation.

\subsection{Quantum Fields in a 5D Expanding Universe}
We briefly summarize quantization of field operators. See Appendix~\ref{quantum field in openFRW} for details including notations. We decompose 5D scalar and spinor fields as
\begin{align}
    \hat{\phi}_i(x,y)=&\frac{1}{a^{\frac32}(t)\sqrt{2\pi R}}\sum_{n\in \mathbb Z}\hat{\phi}^{(i)}_n(x)e^{\ri\frac{ny}{R}},\\
    \hat{\Psi}_I(x,y)=&\sum_{n\in \mathbb Z}\frac
    {b^{\frac14}(t)}{a^{\frac32}(t)}\left(\begin{array}{c}\hat{\psi}_n^{(I)}(x)\\ \hat{\zeta}^{(I)\dagger}_n(x)\end{array}\right)e^{\ri\frac{ny}{R}}.
\end{align}
Each 4D field can be further expanded as
\begin{align}
    \hat{\phi}^{(i)}_{n}(x)=&\int_0^\infty dq\sum_{l=0}^\infty\sum_{m=-l}^l\left[\hat{a}^{(i)}_{nqlm} f^{(i)}_{n,q}(t)Q_{qlm}(\chi,\theta,\varphi)+\hat{a}^{(i)\dagger}_{-nqlm}\left(f^{(i)}_{n,q}(t)\right)^*Q^*_{qlm}(\chi,\theta,\varphi)\right],\\
    \hat{\psi}_n^{(I)}(x)=&\int_0^\infty d\lambda\sum_{S,j,m}\left[\hat{b}^{(I)S}_{n\lambda jm}f^{(I)S}_{n,\lambda}(t)\Theta^S_{\lambda,j,m}(\chi,\theta,\varphi)-\hat{d}^{(I)S\dagger}_{n\lambda j m}\left(g^{(I)S}_{n,\lambda}(t)\right)^*\Theta^S_{\lambda,j,m}(\chi,\theta,\varphi)\right],\\
    \hat{\zeta}_n^{(I)\dagger}(x)=&\int_0^\infty d\lambda\sum_{S,j,m}\left[\hat{b}^{(I)S}_{n\lambda jm}g^{(I)S}_{n,\lambda}(t)\Theta^S_{\lambda,j,m}(\chi,\theta,\varphi)+\hat{d}^{(I)S\dagger}_{n\lambda j m}\left(f^{(I)S}_{n,\lambda}(t)\right)^*\Theta^S_{\lambda,j,m}(\chi,\theta,\varphi)\right],
\end{align}
where the quantum numbers run over $S=\pm$, $j=\frac12,\frac32,\cdots$, $m=-j,\cdots,j$, and the mode functions $Q_{qlm}(\chi,\theta,\varphi)$, $\Theta^S_{\lambda,j,m}(\chi,\theta,\varphi)$ are defined, respectively, in \eqref{defQ} and \eqref{defTheta}. The scalar mode function satisfies
\begin{align}
    \ddot{f}_{n,q}^{(i)}(t)+\omega_{(i)nq}^2(t)f_{n,q}^{(i)}(t)=0,
\end{align}
where the one-particle energy of the $n$-th KK mode is defined as
\begin{align}
    \omega_{(i)nq}^2(t)=\frac{q^2+1}{a^2(t)r_K^2}+\frac{m^2_{i}}{b}+\frac{n^2}{R^2b^3}-\frac94H^2-\frac32\dot{H},
\end{align}
and the mode function is normalized as $f^{(i)}_{n,q}(t)\left(\dot{f}^{(i)}_{n,q}(t)\right)^*-\left(f^{(i)}_{n,q}(t)\right)^*\dot{f}^{(i)}_{n,q}(t)=\ri$. The spinor mode functions satisfy
\begin{align}
     \left(\begin{array}{c}\dot{f}^{(I)S}_{n,\lambda}\\ \dot{g}^{(I)S}_{n,\lambda}\end{array}\right)=&-\ri\left(\begin{array}{cc} \frac{S\lambda}{a r_K}&\bar{M}^{(I)}_n\\ M^{(I)}_n&-\frac{S\lambda}{ar_K}\end{array}\right)\left(\begin{array}{c} f^{(I)S}_{n,\lambda}\\ g^{(I)S}_{n,\lambda}\end{array}\right)\nonumber\\
     =&-\ri\omega_{(I)n\lambda}\left(\begin{array}{cc} \cos\vartheta_{(I)n\lambda S}&e^{-\ri c_{(I)n}}\sin\vartheta_{(I)n\lambda S}\\e^{+\ri c_{(I)n}}\sin\vartheta_{(I)n\lambda S}&-\cos\vartheta_{(I)n\lambda S}\end{array}\right)\left(\begin{array}{c} f^{(I)S}_{n,\lambda}\\ g^{(I)S}_{n,\lambda}\end{array}\right)
\end{align}
where the $n$-th KK mass is
\begin{align}
    M^{(I)}_n=\frac{m_{I}}{b^{\frac12}}-\frac{\ri n}{Rb^{\frac32}}=e^{\ri c_{(I)n}}\sqrt{\frac{m_{I}^2}{b}+\frac{n^2}{R^2b^3}},
\end{align}
the one-particle energy is
\begin{align}
\omega_{(I)n\lambda}(t)=\sqrt{\frac{\lambda^2}{a^2(t)r_K^2}+\left|M_n^{(I)}(t)\right|^2},
\end{align}
and we have introduced $\vartheta_{(I)n\lambda S}$ such that
\begin{align}
    \sin \vartheta_{(I)n\lambda S}=\frac{\left|M_n^{(I)}\right|}{\omega_{(I)n\lambda}}, \qquad \cos\vartheta_{(I)n\lambda S}=\frac{S\lambda}{ar_K\omega_{(I)n\lambda}}.
\end{align}

We need to define the vacuum state that is annihilated by the annihilation operators introduced above. More precisely speaking, the above expansion of operators is yet rather formal unless we specify the mode functions. For scalar fields, we use the following formal WKB solution
\begin{align}
    f_{n,q}^{(i)}(t)=\frac{\alpha_{(i)n,q}(t)}{\sqrt{2\omega_{(i)nq}(t)}}e^{-\ri\int^t dt'\omega_{(i)nq}(t')}+\frac{\beta_{(i)n,q}(t)}{\sqrt{2\omega_{(i)nq}(t)}}e^{+\ri\int^t dt'\omega_{(i)nq}(t')},
\end{align}
where the time-dependent Bogoliubov coefficients satisfy the mode equations
\begin{align}
   \left( \begin{array}{c}\dot{\alpha}_{(i)nq}(t)\\ \dot{\beta}_{(i)nq}(t)\end{array}\right)=\frac{\dot\omega_{(i)nq}(t)}{2\omega_{(i)nq}(t)}\left(\begin{array}{cc}0&e^{+2\ri\int^t dt'\omega_{(i)nq}(t')}\\e^{-2\ri\int^t dt'\omega_{(i)nq}(t')}&0\end{array} \right)\left(\begin{array}{c}\alpha_{(i)n,q}(t)\\ \beta_{(i)n,q}(t)\end{array}\right),
\end{align}
and the normalization condition
\begin{align}
    |\alpha_{(i)n,q}(t)|^2-|\beta_{(i)n,q}(t)|^2=1
\end{align}
which follows from the normalization condition of $f_{n,q}^{(i)}(t)$. For spinors, we introduce base functions
\begin{align}
    {\bm v}_{(I)n\lambda S}^+=\left(\begin{array}{c}e^{-\ri c_{(I)n}}\cos\frac12\vartheta_{(I)n\lambda S}\\ \sin\frac12\vartheta_{(I)n\lambda S}\end{array}\right),\qquad   {\bm v}_{(I)n\lambda S}^-=\left(\begin{array}{c}-e^{-\ri c_{(I)n}}\sin\frac12\vartheta_{(I)n\lambda S}\\ \cos\frac12\vartheta_{(I)n\lambda S}\end{array}\right).
\end{align}
Then, we write the set of spinors' mode functions as
\begin{align}
    \left(\begin{array}{c} f^{(I)S}_{n,\lambda}(t)\\ g^{(I)S}_{n,\lambda}(t)\end{array}\right)=\gamma_{(I)n\lambda S}(t)e^{-\ri \int^t dt'\omega_{(I)n\lambda}(t')}{\bm v}_{(I)n\lambda S}^+(t)+\delta_{(I)n\lambda S}(t)e^{+\ri \int^t dt'\omega_{(I)n\lambda}(t')}{\bm v}_{(I)n\lambda S}^-(t).
\end{align}
The initial conditions at $t=t_{\rm ini}$ are taken to be
\begin{align}
    \alpha_{(i)n,q}(t_{\rm ini})=1,\quad \beta_{(i)n,q}(t_{\rm ini})=0, \quad \gamma_{(I)n\lambda S}(t_{\rm ini})=1,\quad \delta_{(I)n\lambda S}(t_{\rm ini})=0
\end{align}
for all sets of quantum numbers. These conditions completely fix the mode functions and the vacuum state $|0_{\rm in}\rangle$ satisfying
\begin{align}
    \hat{a}^{(i)}_{nqlm}|0_{\rm in}\rangle=0=\hat{b}^{(I)S}_{n\lambda jm}|0_{\rm in}\rangle=\hat{d}^{(I)S}_{n\lambda jm}|0_{\rm in}\rangle.
\end{align}
We note that thus-defined adiabatic vacuum state is not a unique choice since there is no unique vacuum in the absence of a global timelike Killing vector. One could take e.g. the higher-order WKB solution or the exact WKB solutions, but for a technical simplicity, we consider the above (lowest order) adiabatic vacuum state as the basis of our model.

Let us take the initial state as a thermal state, which is characterized by the expectation values of the creation and the annihilation operators as
\begin{align}
    \left\langle\hat{a}_{(i)nqlm}^\dagger\hat{a}_{(i')n'q'l'm'}\right\rangle_\beta=&n_{B}\left(\omega_{(i)nq}(t_{\rm ini})\right)\delta_{ii'}\delta_{nn'}\delta(q-q')\delta_{ll'}\delta_{mm'},\label{cc1}\\
    \left\langle\hat{a}_{(i)nqlm}\hat{a}_{(i')n'q'l'm'}^\dagger\right\rangle_\beta=&\left(n_{B}(\omega_{(i)nq}(t_{\rm ini}))+1\right)\delta_{ii'}\delta_{nn'}\delta(q-q')\delta_{ll'}\delta_{mm'},\label{cc2}
\end{align}
for scalar fields whereas the spinor fields satisfy
\begin{align}
    &\left\langle\hat{b}^{(I)S\dagger}_{n\lambda jm}\hat{b}^{(I')S'}_{n'\lambda'j'm'}\right\rangle_\beta=n_F(\omega_{(I)n\lambda}(t_{\rm ini}))\delta_{II'}\delta_{nn'}\delta_{SS'}\delta(\lambda-\lambda')\delta_{jj'}\delta_{mm'}=\langle\hat{d}^{(I)S\dagger}_{n\lambda jm}\hat{d}^{(I')S'}_{n'\lambda'j'm'}\rangle\label{cac1}\\
   &\left \langle\hat{b}^{(I)S}_{n\lambda jm}\hat{b}^{(I')S'\dagger}_{n'\lambda'j'm'}\right\rangle_\beta=\left\{1-n_F(\omega_{(I)n\lambda}(t_{\rm ini}))\right\}\delta_{II'}\delta_{nn'}\delta_{SS'}\delta(\lambda-\lambda')\delta_{jj'}\delta_{mm'}=\langle\hat{d}^{(I)S}_{n\lambda jm}\hat{d}^{(I')S'\dagger}_{n'\lambda'j'm'}\rangle,\label{cac2}
\end{align}
for spinor fields. Here, the factors $n_{B,F}(\omega)$ are Bose-Einstein (Fermi-Dirac) distributions at the temperature $T=\beta^{-1}$
\begin{align}
    n_{B}(\omega)=\frac{1}{e^{\beta\omega}-1},\qquad n_F(\omega)=\frac{1}{e^{\beta\omega}+1}.
\end{align}

\subsection{Energy-Momentum Tensor}\label{sec:EMtensor}
We evaluate the energy-momentum tensor in the Einstein equation~\eqref{EOM1} and \eqref{EOM2}. Now, it is straightforward to compute the expectation value of the energy-momentum tensor operator. For instance, using~\eqref{cc1},\eqref{cc2}, we find
\begin{align}
    \left\langle\hat{\phi}^2_i(x,y)\right\rangle=&\frac{1}{a^3(2\pi R)}\sum_{n\in \mathbb Z}\int_0^\infty dq\rho_{H^3}(q)\frac{1}{\omega_{(i)nq}(t)}\left\{2n_B(\omega_{(i)nq}(t_{\rm ini}))+1\right\}\left(\frac12+N_{n,q}^{(i)}(t)+R_{n,q}^{(i)}(t)\right),
\end{align}
where $N_{n,q}^{(i)}(t)=|\beta_{(i)n,q}(t)|^2$ and $R_{n,q}^{(i)}(t)\equiv {\rm Re}\left(\alpha_{(i)n,q}(t)\bar\beta_{(i)n,q}(t)e^{-2\ri\int^tdt'\omega_{(i)nq}}\right)$. In deriving the above expression, we have used the spectral density $\rho_{H^3}(q)$ defined in \eqref{scalar spectral}. We can decompose the expectation value into three parts as
\begin{align}
    \left\langle\hat{\phi}^2_i(x,y)\right\rangle_{\rm vac}\equiv& \frac{1}{a^3(2\pi R)}\sum_{n\in \mathbb Z}\int_0^\infty dq\rho_{H^3}(q)\frac{1}{2\omega_{(i)nq}(t)},\\
    \left\langle\hat{\phi}^2_i(x,y)\right\rangle_{\rm th}\equiv&\frac{1}{a^3(2\pi R)}\sum_{n\in \mathbb Z}\int_0^\infty dq\rho_{H^3}(q)\frac{1}{\omega_{(i)nq}(t)}n_B(\omega_{(i)nq}(t_{\rm ini})),\\
    \left\langle\hat{\phi}^2_i(x,y)\right\rangle_{\rm non}\equiv&\frac{1}{a^3(2\pi R)}\sum_{n\in \mathbb Z}\int_0^\infty dq\rho_{H^3}(q)\frac{2n_B(\omega_{(i)nq}(t_{\rm ini}))+1}{\omega_{(i)nq}(t)}\left(N_{n,q}^{(i)}(t)+R_{n,q}^{(i)}(t)\right).
\end{align}
The first one corresponds to the (adiabatic) vacuum fluctuation in the time-dependent background. The second one is the initial thermal excitation. The third one becomes non-vanishing only when $|\beta_{(i)n,q}(t)|\neq0$, or equivalently only when the particle is created from vacuum. The first two are present independently of the dynamics of the scalar field whereas the third one becomes important only when the corresponding particles are produced. Note also that the UV divergent contribution appears only from the first one since both the thermal spectrum $n_B$ and the particle production effect $N^{(i)}_{n,q}$ and $R_{n,q}^{(i)}$ are exponentially suppressed for high energy modes in smooth backgrounds. We now give the explicit formulas for the quantities that appear in \eqref{EOM1} and \eqref{EOM2}:
The energy density of the scalar fields is\footnote{In deriving the scalar energy density, one needs to use $\lim_{\bm x\to\bm x'}{\bm \nabla}_{\bm x}\cdot{\bm \nabla}_{\bm x'}\Pi_q(\bm x,\bm x')=(q^2+1)\rho_{H^3}(q)$, where ${\bm \nabla}_{\bm x}$ is understood as the covariant derivative on $H^3$ with respect to $\bm x$ and the contraction includes the metric on $H^3$.}
\begin{align}
    \langle\hat\rho_\phi\rangle_{\beta}=&\frac{1}{a^3(2\pi R)}\sum_{i}\sum_{n\in\mathbb Z}\int_0^\infty dq\rho_{H^3}(q)\frac{2n_B(\omega_{(i)nq}(t_{\rm ini}))+1}{2\omega_{(i)nq}(t)}\nonumber\\
    &\times\Biggl[\left(\omega_{(i)nq}^2+\frac34\dot H\right)\left(1+2N^{(i)}_{n,q}(t)\right)+\left(\frac{9H^2}{2}+\frac{3\dot H}{2}\right)R^{(i)}_{n,q}(t)-6H\omega_{(i)nq}I^{(i)}_{n,q}(t)\Biggr],
\end{align}
where $I^{(i)}_{n,q}(t)={\rm Im}\left(\alpha_{(i)n,q}(t)\bar\beta_{(i)n,q}(t)e^{-2\ri\int^tdt'\omega_{(i)nq}}\right)$. The energy density of spinor fields is
\begin{align}
    \langle\hat{\rho}_{\Psi}\rangle_{\beta}=&\frac{1}{a^3(2\pi R)}\sum_{I}\sum_{n\in\mathbb Z}\sum_{S=\pm}\int_0^\infty d\lambda\frac{\rho_{H^3}^f(\lambda)}{2}\omega_{(I)n\lambda}(t)\left(2n_{F}(\omega_{(I)n\lambda}(t_{\rm ini}))-1\right)\left(1-2N^{(I)}_{n,\lambda,S}(t)\right),
\end{align}
where $N^{(I)}_{n,\lambda,S}(t)=\left|\delta_{(I)n\lambda S}\right|^2$ is the number density of spin $1/2$ particles produced from vacuum. The source term of the radion from scalar fields is
\begin{align}
    \langle\hat{J}_\phi\rangle_\beta=&\frac{b}{a^3(2\pi R)}\sum_i\sum_{n\in\mathbb Z}\int_0^\infty dq \rho_{H^3}\frac{2n_B(\omega_{(i)nq}(t_{\rm ini}))+1}{2\omega_{(i)nq}(t)}\left(\frac{n^2}{b^3R^2}+\frac{m_i^2}{3b}\right)\left(1+2N^{(i)}_{n,q}(t)+2R^{(i)}_{n,q}(t)\right),
\end{align}
and that of spinor fields is
\begin{align}
    \langle \hat{J}_\Psi\rangle_\beta=&\frac{b}{a^3(2\pi R)}\sum_I\sum_{n\in\mathbb Z}\sum_{S=\pm}\int_0^\infty d\lambda \frac{\rho_{H^3}^f(\lambda)}{2}\left(2n_{F}(\omega_{(I)n\lambda}(t_{\rm ini}))-1\right)\nonumber\\
    &\times\Biggl[\frac{1}{\omega^{(I)}_{n\lambda}(t)}\left(\frac{m_I^2}{3b(t)}+\frac{n^2}{R^2b^3(t)}\right)\left(1-2N^{(I)}_{n,\lambda,S}(t)\right)\nonumber\\
    &\quad +\frac{2S\lambda}{\omega_{(I)n\lambda}(t)a(t)r_K\left|M_n^{(I)}(t)\right|}\left(\frac{m_I^2}{3b(t)}+\frac{n^2}{R^2b^3(t)}\right)R^{(I)}_{n,\lambda,S}(t)+\frac{4m_In}{3|M_n^{(I)}(t)|Rb^2(t)}I^{(I)}_{n,\lambda,S}(t)\Biggr],
\end{align}
where we have introduced $\gamma_{(I)n\lambda S}(t)\bar{\delta}_{(I)n\lambda S}(t)e^{-2\ri \int^t dt'\omega_{(I)n\lambda}(t')}=R^{(I)}_{n,\lambda,S}(t)+\ri I^{(I)}_{n,\lambda,S}(t)$ where $R^{(I)}_{n,\lambda,S}(t),I^{(I)}_{n,\lambda,S}(t)\in \mathbb R$. Thus, we are able to solve the dynamics of the universe and matter fields in a consistent manner, at least in principle. 

In the following discussion, we consider only the effects of initial thermal effects and neglect particle production effects due to the time-dependent background. Such approximation is justified as long as the Hubble scale at the time of modulus stabilization $H(t_{\rm mod})$ is much less than the KK scale at that time $m_{\rm KK}(t_{\rm mod})$. Typically, particle production from vacuum occurs when the time dependent energy $\omega_k(t)$ of a particle becomes locally minimized. Within our setup, KK particle masses are minimized at the time of radion stabilization $t=t_{\rm mod}$. The time scale of KK particle energy is governed by the Hubble scale, which determines the KK particle production ratio. However, since $H(t_{\rm mod})\ll m_{\rm KK}(t_{\rm mod})$, we do not expect gravitational particle production, and therefore, the particle production effects due to the time dependent background can be neglected. Thus, the energy density and source term can be significantly simplified as
\begin{align}
     \langle\hat\rho_\phi\rangle_{\beta}\approx& \frac{1}{a^3(t)(2\pi R)}\sum_{i}\sum_{n\in\mathbb Z}\int_0^\infty dq\rho_{H^3}(q)\frac{2n_B(\omega_{(i)nq}(t_{\rm ini}))+1}{2\omega_{(i)nq}(t)}\left(\omega_{(i)nq}^2(t)+\frac34\dot H(t)\right),\label{rhophi}\\
     \langle\hat{\rho}_{\Psi}\rangle_{\beta}\approx&\frac{1}{a^3(t)(2\pi R)}\sum_{I}\sum_{n\in\mathbb Z}\int_0^\infty d\lambda\rho_{H^3}^f(\lambda)\omega_{(I)n\lambda}(t)\left(2n_{F}(\omega_{(I)n\lambda}(t_{\rm ini}))-1\right),\label{rhopsi}\\
     \langle\hat{J}_\phi\rangle_\beta\approx&\frac{b(t)}{a^3(t)(2\pi R)}\sum_i\sum_{n\in\mathbb Z}\int_0^\infty dq \rho_{H^3}\frac{2n_B(\omega_{(i)nq}(t_{\rm ini}))+1}{2\omega_{(i)nq}(t)}\left(\frac{n^2}{b^3(t)R^2}+\frac{m_i^2}{3b(t)}\right),\label{Jphi}\\
      \langle \hat{J}_\Psi\rangle_\beta\approx &\frac{b(t)}{a^3(t)(2\pi R)}\sum_I\sum_{n\in\mathbb Z}\int_0^\infty d\lambda \rho_{H^3}^f(\lambda)\frac{2n_{F}(\omega_{(I)n\lambda}(t_{\rm ini}))-1}{\omega_{(I)n\lambda}(t)}\left(\frac{n^2}{R^2b^3(t)}+\frac{m_I^2}{3b(t)}\right).\label{Jpsi}
\end{align}
As noted above, the terms proportional to $n_{B,F}$ exponentially decay at large $q,\lambda,n$, namely high momentum region, and therefore, we do not expect UV divergences. For the vacuum fluctuation part, we apply dimensional regularization as follows. Let us consider the vacuum part of the scalar energy density
\begin{align}
    \langle\hat{\rho}_\phi\rangle_0
    =&\frac{1}{2\pi R}\sum_i\sum_{n\in\mathbb Z}\int_0^\infty \frac{d\kappa \kappa^2}{2\pi^2}\left(\frac12\omega_{(i)nq}+\frac{3\dot H}{8\omega_{(i)nq}}\right),
\end{align}
where we have changed the integration variable $\kappa=\frac{q}{ar_K}$. Now the one particle energy is
\begin{align}
    \omega_{(i)nq}=\sqrt{\kappa^2+\frac{1}{a^2r_K^2}+\frac{m^2_{i}}{b}+\frac{n^2}{R^2b^3}-\frac94H^2-\frac32\dot{H}}.
\end{align}
We now apply the dimensional regularization by deforming the integral as
\begin{align}
   \langle\hat{\rho}_\phi\rangle_0\to  \frac{\mu^{2\epsilon}}{2\pi R}\sum_i\sum_{n\in\mathbb Z}\int_0^\infty \frac{d\kappa \kappa^{2-2\epsilon}}{2\pi^2}\left(\frac12\omega_{(i)nq}+\frac{3\dot H}{8\omega_{(i)nq}}\right),
\end{align}
where $\mu$ is a renormalization scale. Performing integration, one finds
\begin{align}
    \langle\hat{\rho}_\phi\rangle_0=\frac{\mu^{2\epsilon}}{2\pi R}\sum_i\sum_{n\in \mathbb Z}\frac{\Gamma\left(\frac32-\epsilon\right)\Gamma\left(-1+\epsilon\right)}{16\pi^{\frac52}}\left[\frac{C_{(i)n}^{2-\epsilon}}{2-\epsilon}+\frac{3\dot{H}C_{(i)n}^{1-\epsilon}}{2}\right],\label{dim reg}
\end{align}
where $C_ {(i)n}\equiv \frac{1}{a^2r_K^2}+\frac{m^2_{i}}{b}+\frac{n^2}{R^2b^3}-\frac94H^2-\frac32\dot{H}$. We use the following analytic continuation formula~\cite{Kirsten:2010zp},
\begin{align}
\sum_{n\in\mathbb Z}\left(A+Bn^2\right)^{-s}=\frac{\sqrt\pi}{\Gamma(s)\sqrt B}\left[\Gamma\left(s-\frac12\right)A^{\frac12-s}+4\sum_{n=1}^\infty\left(\frac{\pi^2n^2}{AB}\right)^{\frac12(s-\frac12)}K_{\frac12-s}\left(2\pi n\sqrt{\frac AB} \right)\right]
\end{align}
for $A,B>0$.\footnote{This formula should be applied with care. The effective frequency square of scalar fields may become negative due to the curvature terms. As discussed in Appendix~\ref{appD}, the tachyonic instability would not appear in our setup with an appropriate choice of parameters. However, in general, one cannot use the above formula when there is an unstable mode leading to $A<0$.} Then the vacuum energy can be evaluated as
\begin{align}
    \langle\hat{\rho}_\phi\rangle_0=&\sum_i\frac{m_{(i)0}^5b^{\frac32}}{8\pi^2}\Biggl[\left(\frac{1}{15}+2\sum_{n=1}^\infty e^{-x_{(i)n}}\frac{3+3x_{(i)n}+x_{(i)n}^2}{x_{(i)n}^5}\right)+\frac{\dot H}{4m_{(i)0}^2}\left(1+6\sum_{n=1}^\infty e^{-x_{(i)n}}\frac{1+x_{(i)n}}{x_{(i)n}^3}\right)\Biggr],
\end{align}
where 
\begin{align}
    m_{(i)0}=&\sqrt{\frac{1}{a^2r_K^2}+\frac{m^2_i}{b}-\frac94H^2-\frac32\dot{H}},\\
    x_{(i)n}=&2\pi n m_{(i)0}Rb^{\frac32}.
\end{align}
Note that we have taken $\epsilon\to0$ but no UV divergent terms appear within the dimensional regularization.\footnote{We note that the background fields $a(t),b(t)$ appear in expectation values of physical quantities as a consequence of the curved background. If there appear UV divergences, we need to add counterterms such as $S_{\rm ct}=\int dx^5\sqrt{-g_5}\left[\alpha_1+\alpha_2{\cal R}_5+\alpha_3{\cal R}^2+\alpha_4{\cal R}_{MN}{\cal R}^{MN}+\cdots\right]$, where $\alpha_i$ are divergent constants. For instance, by substituting background, the bulk cosmological constant $\alpha_1$ gives $S=\int dt d^3\bm x a^3\frac{2\pi R \alpha_1}{b}$, which can be the counterterm for radion potential. The local UV-sensitive pieces are therefore understood to be absorbed into the renormalized coefficients of the five-dimensional effective action. We emphasize, however, that in the dimensional-regularization prescription used in this work, the physical quantities that we need to evaluate contain no remaining UV pole after the full five-dimensional contribution is included.}  In the same way, we obtain
\begin{align}
    \langle\hat\rho_{\Psi}\rangle_0=&-\sum_I\frac{m_I^5}{2\pi^2b}\Biggl[\left(\frac{1}{15}+2\sum_{n=1}^\infty \frac{e^{-x_{(I)n}}\left(3+3x_{(I)n}+x_{(I)n}^2\right)}{x_{(I)n}^5}\right)\nonumber\\
    &\hspace{2cm}-\frac{1}{4a^2r_K^2m_I^2}\left(\frac13+2\sum_{n=1}^\infty \frac{e^{-x_{(I)n}}\left(1+x_{(I)n}\right)}{x_{(I)n}^3}\right)\Biggr],
\end{align}
where 
\begin{align}
    x_{(I)n}=2\pi n m_IRb.
\end{align}
Also, we obtain
\begin{align}
    \langle\hat{J}_{\phi}\rangle_0=&\frac{b^{\frac52}}{8\pi^2}\sum_{i}\Biggl[-\frac{m_{(i)0}^5}{15}+2m_{(i)0}^5\sum_{n=1}^\infty \frac{e^{-x_{(i)n}}\left(12+12x_{(i)n}+3x_{(i)n}^2-x_{(i)n}^3\right)}{x_{(i)n}^5}\nonumber\\
    &\hspace{3.5cm} +\frac{m_i^2m_{(i)0}^3}{9b}+\frac{2m_i^2m_{(i)0}^3}{3b}\sum_{n=1}^\infty\frac{e^{-x_{(i)n}}\left(1+x_{(i)n}\right)}{x_{(i)n}^3}\Biggr],
\end{align}
\begin{align}
    \langle \hat{J}_{\Psi}\rangle_0=&-\sum_I\frac{m_I^5}{\pi^2}\Biggl[\frac{1}{45}+2\sum_{n=1}^\infty \frac{e^{-x_{(I)n}}\left(18+18x_{(I)n}+5x_{(I)n}^2-x_{(I)n}^3\right)}{3x_{(I)n}^5}\nonumber\\
    &\hspace{4.cm}-\frac{b}{6a^2r_K^2m_I^2}\sum_{n=1}^\infty \frac{e^{-x_{(I)n}}\left(3+3x_{(I)n}+x_{(I)n}^2\right)}{x_{(I)n}^3}\Biggr].
\end{align}

Let us show the thermal contribution part. Thermal contributions have no UV divergences but computation of them is even more complicated, which is shown in Appendix~\ref{appB}. Here, we summarize the resulting expressions
\begin{align}
    \langle\hat{\rho}_\phi\rangle_\beta|_{\rm th}\approx& \frac{c_1N_BT^5}{\pi^3a^3b^{\frac32}}-\frac{9c_2N_BT^3H^2b^{\frac32}}{16\pi^3a^3}+\frac{c_2T^3b^{\frac12}}{4\pi^3a^3}\left(\sum_im_i^2\right),\\
    \langle\hat\rho_\Psi\rangle_\beta|_{\rm th}\approx&\frac{4c_1N_FT^5}{\pi^3a^3b^{\frac32}}+\frac{3c_3T^3}{4\pi^3a^3b^{\frac32}r_K^2}+\frac{c_2T^3b^{\frac12}}{\pi^3a^3}\left(\sum_Im_I^2\right),\\
    \langle\hat{J}_\phi\rangle_\beta|_{\rm th}\approx& \frac{c_1N_BT^5}{\pi^3a^3b^{\frac12}}+\frac{c_2N_BT^3b^{\frac52}}{\pi^3a^3}\left(\frac{9H^2}{16}+\frac{3\dot H}{8}\right)-\frac{c_2T^3b^{\frac32}}{12\pi^3a^3}\left(\sum_im_i^2\right),\\
    \langle \hat{J}_\Psi\rangle_\beta|_{\rm th}\approx&\frac{4c_1N_FT^5}{\pi^3a^3b^{\frac12}}+\frac{3c_3T^3}{4\pi^3a^3b^{\frac12 }r_K^2}-\frac{c_2T^3b^{\frac32}}{3\pi^3 a^3}\left(\sum_Im_I^2\right),
\end{align}
where $c_1=\frac{15}{e}+\frac34\approx6.3$, $c_2= \frac4e +1\approx 2.5$, $c_3=\frac2e+\frac14\approx 1.0$. These expressions are the leading order in the high-temperature expansion in $T^{-1}$, and we have dropped the terms including $\gamma=b^3/a^2$ as they are small corrections, which we discuss in Appendix~\ref{appB}.

We give a few comments about the thermal corrections:
\begin{itemize}
    \item One may expect that, as the volume of extra dimension increases, more KK modes contribute to the thermal energy, which may lead to the effective potential with positive powers of $b$. The positive powers of $b$ appearing below arise only as subleading terms in the expansion of the one-particle energy, not from an increase in the number of active KK modes. In our model, we have assumed the initial state to be a thermal equilibrium state with the temperature $T=\beta^{-1}$ but the fields evolve as free particles, which describes the collisionless approximation for particles. Accordingly, the KK modes contributing to the energy density are fixed by the initial temperature. Therefore, the total number of active KK modes never increases and the energies of the already populated KK modes redshift according to the time-dependent background. This is somewhat different from the expectation from the equilibrium state analysis in the static spacetime~\cite{Dienes:1998hx}.
    \item We should also emphasize that the numerical factors $c_{1,2,3}$ appear as a consequence of our (crude) approximation in computing the thermal contributions. 
    \item The overall scaling can be understood as follows. The initial
five-dimensional phase-space density gives a factor of order $T^4$. Since
we work in the collisionless approximation, this number of excited modes is
fixed by the initial thermal state and is diluted only by the expansion of
the non-compact three-dimensional space, giving a factor $a^{-3}$. In the
regime where the one-particle energy is dominated by the KK momentum,
$E_{\rm KK}\sim T/b^{3/2}$, the leading thermal energy density scales as
$$\rho_{\rm th}\sim T^4\times a^{-3}\times \frac{T}{b^{3/2}}=\frac{T^5}{a^3b^{3/2}} .$$
    \item The positive powers of $b$ appearing in the bulk-mass terms do not
represent the true large-$b$ behavior. They arise as subleading corrections
in the expansion around the KK-momentum dominated energy. Schematically,
$$\omega_n=\sqrt{\frac{n^2}{R^2b^3}+\frac{m^2}{b}+\cdots}=\frac{|n|}{Rb^{3/2}}\left(1+\frac{m^2R^2b^2}{2n^2}+\cdots\right).$$
For thermally populated KK modes, $n\sim RT$, and hence the relative size of the mass correction is of order $(mb/T)^2$. Therefore the mass-dependent terms with positive powers of $b$ should not be extrapolated to large $b$;
the expansion is reliable only while the KK momentum dominates the one-particle energy. The same caution applies to the terms proportional to the curvature and Hubble-induced quantities, such as $H^2$, $\dot H$, and $r_K^{-2}$. 
\end{itemize} 

The thermal energy at the leading order is indeed problematic for radion stabilization. Indeed, we expect that the radion should be stabilized by the vacuum contribution as 3D space expansion would eventually dilute the thermal potential. More specifically, we would consider the radion stabilization in the late universe is achieved by
\begin{align}
    V_{\rm eff}=\Lambda_{4D}e^{-\sqrt{\frac23} \frac{\xi}{M_{\rm pl}}}+2\pi R(\langle\hat\rho_\phi\rangle_0|_{H\to0,a\to\infty}+\langle\hat\rho_\Psi\rangle_0|_{a\to\infty}),
\end{align}
where $\cdot|_{H\to0,a\to\infty}$ implies that the thermal, Hubble-induced and spatial-curvature-induced corrections disappear. When the thermal corrections exceed $V_{\rm eff}$, the local minimum of $V_{\rm eff}$ disappears, and the radion is destabilized. Similar effects have been discussed in the literature~\cite{Buchmuller:2004tz,Kallosh:2004yh,Barreiro:2007hb}. Then, one of the important questions is whether the thermal contributions are diluted before the radion reaches the local minimum. This makes the stabilization of the radion even more difficult.

In the next subsection, we examine whether the radion stabilization can be achieved starting from the expanding phase of both the 3D non-compact space and the extra compact dimension.

\subsection{Dynamical Compactification and the Emergence of a 4D Universe}\label{sec:4Demergence}
In this subsection, we solve the coupled background-radion dynamics and determine under what conditions an initially expanding five-dimensional universe can dynamically evolve into an effectively four-dimensional one.

First, we focus on the vacuum contribution to the effective potential and look for the radion potential local minimum at which the size of the compact space in the late universe is fixed. Let us recall that the radion potential due to the bulk matter and the cosmological constant is given by
\begin{align}
    V_{\rm eff}=&\frac{\Lambda_{4D}}{b}+\sum_i\frac{R m_{i}^5}{4\pi b}\left(\frac{1}{15}+2\sum_{n=1}^\infty \frac{e^{-\tilde{x}_{(i)n}}\left(3+3\tilde{x}_{(i)n}+\tilde{x}_{(i)n}^2\right)}{\tilde{x}_{(i)n}^5}\right)\nonumber\\
    &-\sum_I\frac{Rm_I^5}{\pi b}\left(\frac{1}{15}+2\sum_{n=1}^\infty \frac{e^{-x_{(I)n}}\left(3+3x_{(I)n}+x_{(I)n}^2\right)}{x_{(I)n}^5}\right),\label{Veff vac}
\end{align}
where $\tilde{x}_{(i)n}=2\pi n m_i Rb$, which corresponds to $x_{(i)n}|_{H\to 0}$. Here and hereafter, we normalize the initial value of the scale factor as
\begin{align}
    a(t_{\rm ini})=b(t_{\rm ini})=1
\end{align}
where $t_{\rm ini}$ denotes the initial time. In the following, we consider a relatively simple setup with the following parameter set,
\begin{align}
&M_{\rm pl}=1,\ R=10, \ T=0.15,\ \Lambda_{\rm 4D}=-1.261073\times10^{-10}\nonumber\\
    &N_B=100, \ N_F=5,\  m_i=m_B= 7.506550\times 10^{-3} \ (\forall i=1,\cdots,N_B),\nonumber\\
&m_I=m_F=1.261962\times 10^{-3} \ (\forall I=1,\cdots,N_F).
\end{align}
Here the parameters $\Lambda_{\rm 4D},m_{B,F}$ are fine-tuned such that the local minimum of the radion potential \eqref{Veff vac} appears around
\begin{align}
    b_0=10
\end{align}
and the resulting positive value of the potential becomes much smaller than the potential barrier scale, which is essentially the same as the fine-tuning to realize the present cosmological constant scale.\footnote{In the numerical simulation, we add small constant to make the value of the potential energy at $b_0$ to be much smaller. This could be achieved by further tuning both the value of $b_0$ and $\Lambda^5$ or any additional contributions. However, we simply put an adjusting parameter here. Such a small constant does not affect the dynamics of the radion.} Although the parameter choice is only marginally consistent with the thermal expansion, the leading thermal expression is sufficient for our qualitative purpose of estimating the radion-dependent contribution from the initially excited KK population. Note also that the initial Hubble scale is given by $H_{\rm ini}\sim \sqrt{\frac{2c_1}{3\pi^2}(N_B+4N_F)R T^5}\sim 0.19$, which is slightly larger than the temperature. Nevertheless, the Hubble scale soon decreases as the universe expands, whereas \(T\) here simply labels the initial thermal distribution in the collisionless approximation. This justifies neglecting the Hubble-induced terms after the initial transient.\footnote{We also note that since the initial Hubble is large, the initial scalar mass can be approximated by the Hubble-induced terms. From Appendix~\ref{appD}, we find $\mu^2_B\approx \frac{3w\rho}{4M_{\rm pl}^2}$ where $\rho$ is the energy density and $w$ is the equation of state parameter. Now, at the initial time, we assume the thermal energy is dominant, which is proportional to $a^{-3}b^{-\frac32}$. We define the effective equation of state parameter by $\dot\rho=-3H(1+w_{\rm eff}(t))\rho$. Then, we find $w_{\rm eff}=\frac{\dot b}{2Hb}=\frac{\dot\xi}{\sqrt6 HM_{\rm pl}}$, which is exactly the square root of the ratio between the kinetic energy of the radion and the total energy. Then, $w_{\rm eff}$ is very small at the beginning. Therefore, the approximations we use in Appendix~\ref{appB} are satisfied.} In our parameter choice, the parameter relation discussed in Appendix~\ref{appC} is marginally satisfied.

\begin{figure}
    \centering
    \includegraphics[width=0.9\linewidth]{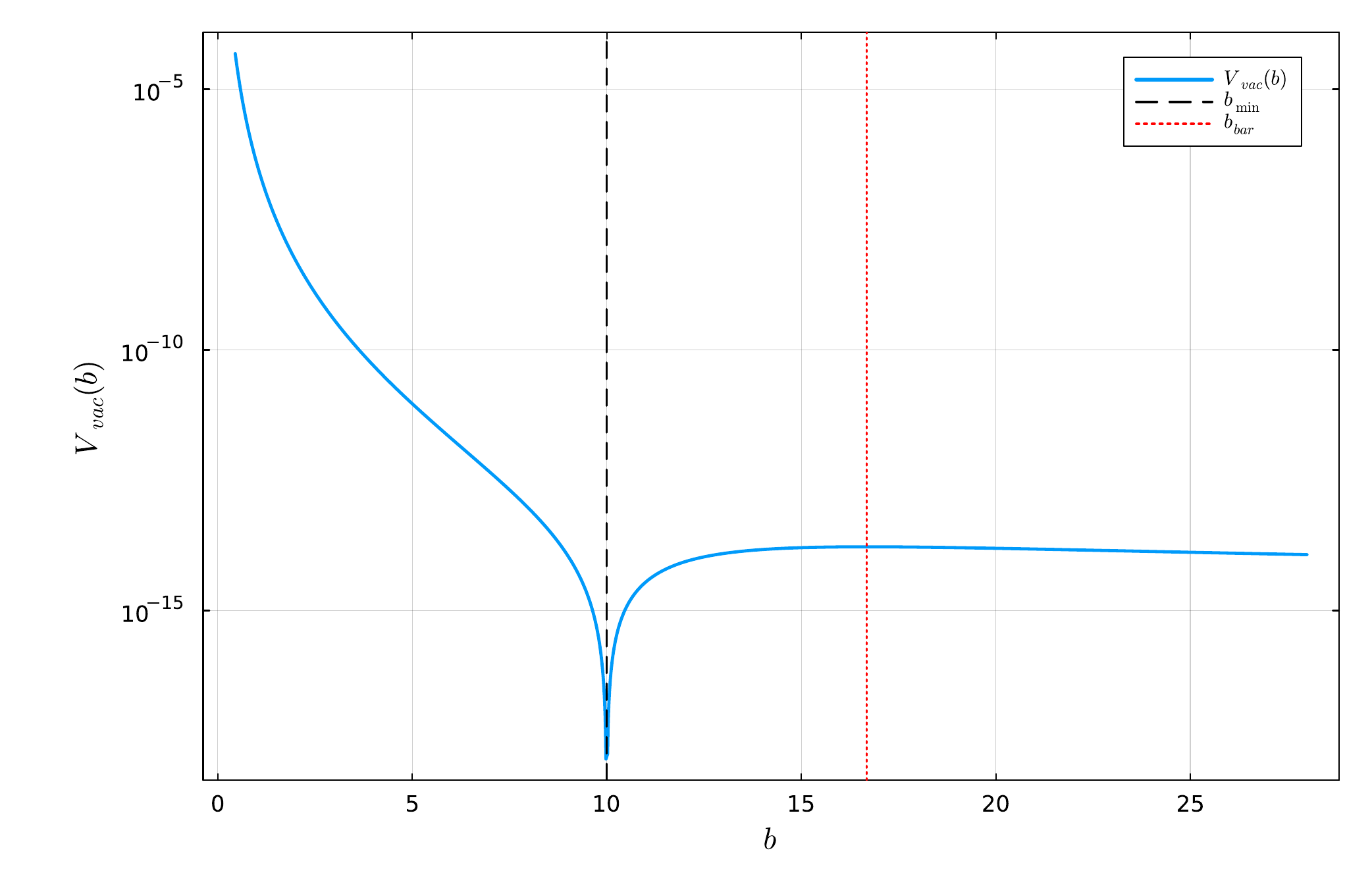}
    \caption{The vacuum effective potential~\eqref{Veff vac}. The vertical axis is log-scaled. Note that the radion dynamics in our scenario starts from $b=1$ at which the potential energy is hierarchically larger than the minimum or the barrier, which naively causes the overshooting problem unless the tracker solution is realized. The local maximum of the potential, namely the barrier, is located at $b=b_{\rm bar}$ indicated by the red dotted line.
    }
    \label{fig:vacuum potential}
\end{figure}

We now introduce the inflaton field in our model. For simplicity, we consider a bulk inflaton field with the $\alpha$-attractor~\cite{Kallosh:2013yoa} type potential
\begin{align}
S=\int d^5x\sqrt{-g_5}\left[-\frac12\partial_M\Phi\partial\Phi^M-\tilde V_0\left(1-e^{-\sqrt{\frac{2}{3\tilde\alpha}}\Phi}\right)\right],
\end{align}
where $\alpha$ and $V_0$ are constant. We identify the zero mode of $\Phi$ as the inflaton, and the zero mode action becomes
\begin{align}
    S=\int dtd\chi d\Omega\sqrt{g_3}a^3\left[\frac12\dot\Phi_0^2-\frac{V_0}{b}\left(1-e^{-\sqrt{\frac{2}{3 \alpha}}\Phi_0}\right)\right],
\end{align}
where $g_3$ is the volume element of $H^3$, $V_0=2\pi R \tilde{V}_0$ and $\alpha=2\pi R\tilde\alpha$. Here we are concerned with the zero mode background field, but including its fluctuations would be possible but would make the discussion more involved, and therefore, we consider the classical part. The potential couples to the radion, and therefore, if the inflation scale is greater than the radion potential barrier scale, radion stabilization would not be achieved, which is a well-known issue~\cite{Kallosh:2004yh}. We emphasize that it is not the problem in our case as we assume the inflation scale is smaller than the radion barrier scale. Nevertheless, one should take into account this fact to realize the inflation consistently with the radion stabilization. After adding the inflaton contribution, we approximate the Einstein equation and the equation of motion of the inflaton as
\begin{align}
   &3H^2M_{\rm pl}^2=\frac12\dot\xi^2+\frac{\Lambda_{\rm 4D}+V_{\rm inf}}{b}+\frac{3M_{\rm pl}^2}{a^2r_K^2}+U_{\rm eff}+\frac12\dot\Phi_0^2\label{EOMr1}\\
   &\ddot\xi+3H\dot\xi-\sqrt{\frac23}\frac{\Lambda_{\rm 4D}+V_{\rm inf}}{M_{\rm pl}b}+\sqrt{\frac23}\frac{2\pi R b}{M_{\rm pl}}\partial_bU_{\rm eff}=0\label{EOMr2}\\
   &\ddot\Phi_0+3H\dot\Phi_0+b^{-1}\partial_{\Phi_0}V_{\rm inf}=0,\label{EOMr3}
\end{align}
where $V_{\rm inf}=V_0\left(1-e^{-\sqrt{\frac{2}{3 \alpha}}\Phi_0}\right)$.
Here, we have approximated the energy density of matter fields to be the leading thermal correction and the vacuum term without Hubble-induced corrections,
\begin{align}
    U_{\rm eff}=&\frac{c_1(N_B+4N_F)T^5}{\pi^3a^3b^{\frac32}}+\frac{N_B\mu_{B}^5b^{\frac32}}{8\pi^2}\left(\frac{1}{15}+2\sum_{n=1}^\infty e^{-x_{Bn}}\frac{3+3x_{Bn}+x_{Bn}^2}{x_{Bn}^5}\right)\nonumber\\
    &-\frac{N_Fm_{F}^5}{2\pi^2b}\left(\frac{1}{15}+2\sum_{n=1}^\infty e^{-x_{Fn}}\frac{3+3x_{Fn}+x_{Fn}^2}{x_{Fn}^5}\right),\label{Ueff}
\end{align}
where $\mu_B=\sqrt{\frac{m_B^2}{b}+\frac{1}{a^2r_K^2}-\frac94H^2-\frac32\dot{H}}$, $x_{Bn}=2\pi n \mu_B R b^{\frac32}$, $x_{Fn}=2\pi n m_FRb$. The curvature and the Hubble-induced corrections are subleading and neglected.\footnote{The initial value of the curvature and the Hubble-induced terms may become greater than the vacuum contributions, especially for the fermionic contribution. However, as the initial phase is dominated by the thermal contribution, the Hubble and the curvature-induced terms are subdominant. Later, the curvature terms are redshifted and therefore, typically the contribution is negligible. If the universe is originally dominated by the curvature or its initial energy density is large enough, the neglected terms become important at the initial stage. However, as the tracker solution is realized eventually, the curvature and the Hubble-induced corrections become smaller.} As expected, a thermal effective potential appears, which will be redshifted by the expansion of both the non-compact and compact spaces. We show the vacuum effective potential $T=0$ in Fig.~\ref{fig:vacuum potential}.

\begin{figure}
    \centering
    \includegraphics[width=0.8\linewidth]{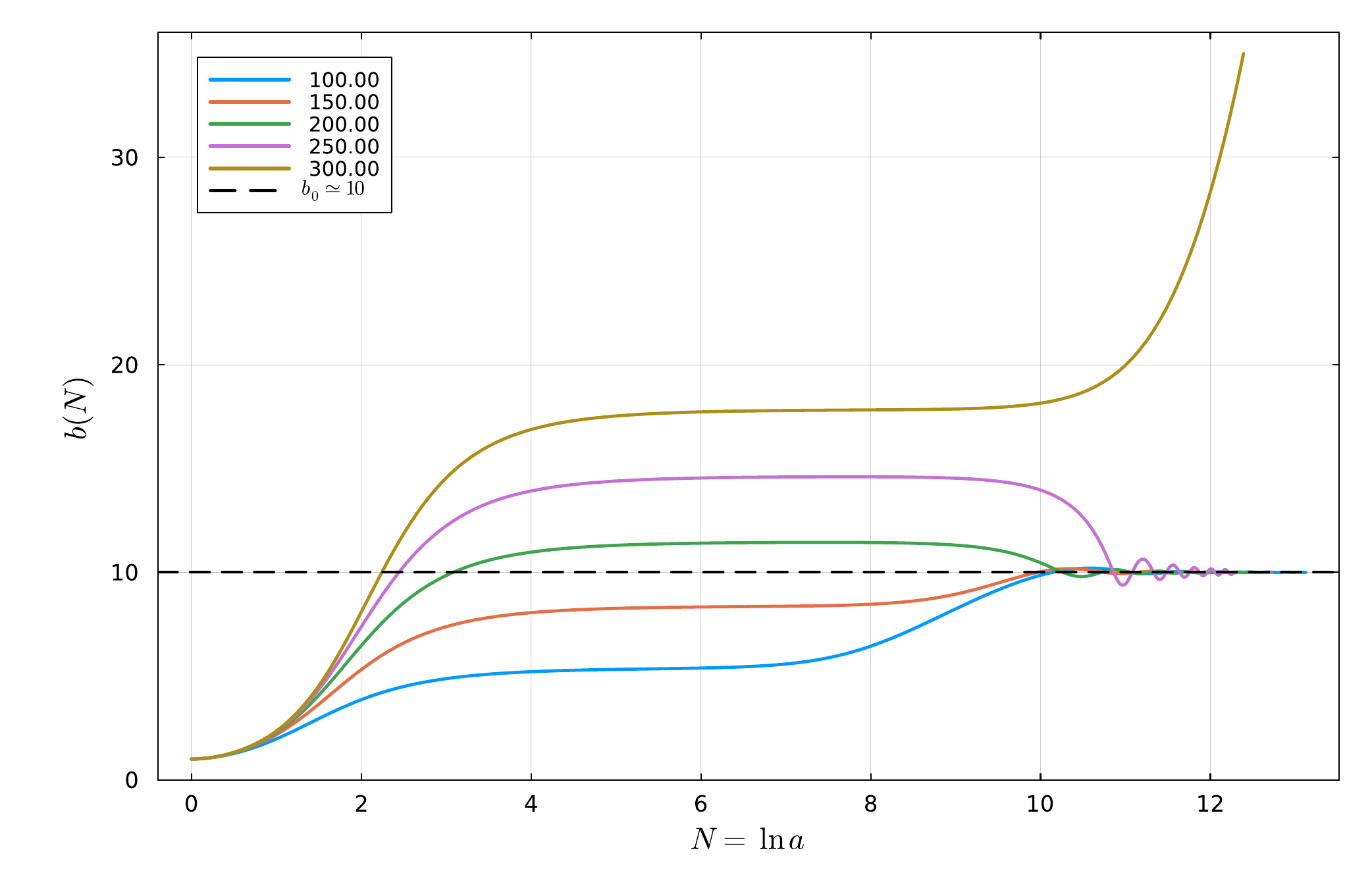}
    \caption{The time-evolution of the radion field without inflation. The labels of different colored lines represent the values of the curvature radius $r_K$. For $r_K=300$, the radion overshoots and decompactification occurs. The dashed black line shows the place of the local minimum.}
    \label{fig:radion tracker wo inflation}
\end{figure}

\begin{figure}
    \centering
    \includegraphics[width=0.7\linewidth]{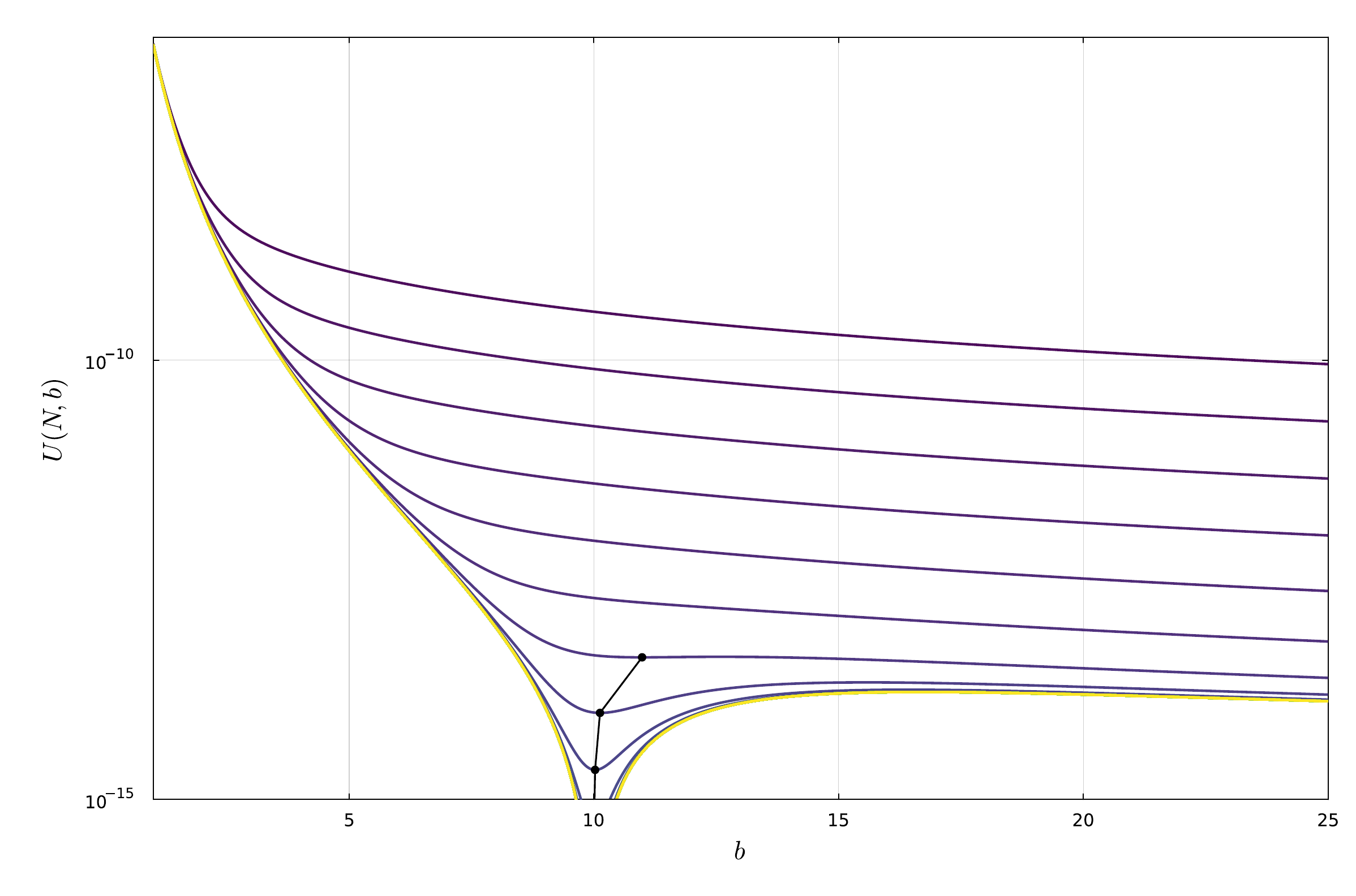}
    \caption{The snapshot of the time-dependent effective potential of the radion~\eqref{Ueff}. The lines from top to bottom correspond to successive shapes of the effective potential. The black dot on each line corresponds to the local minimum. Until the thermal contribution decreases enough, the radion potential does not have a local minimum.  }
    \label{fig:time dependent potential}
\end{figure}

First, we examine the realization of the tracker solution without the inflation stage. We numerically solve the Einstein equation with thermal contributions from bulk matter fields. The dynamics of the radion is shown in Fig.~\ref{fig:radion tracker wo inflation}. The thermal contributions prevent realization of the tracker solution in two ways: First, the thermal contribution defined as
\begin{align}
    U_{\rm th}=\frac{c_1(N_B+4N_F)T^5}{\pi^3a^3b^{\frac32}}
\end{align}
is too shallow to satisfy \eqref{trackercond} even if the curvature dominates the universe. Secondly, while the thermal contribution dominates the universe, the Hubble friction is not strong enough to slow the radion. Furthermore, the thermal correction hides the local minimum of $V_{\rm eff}$ until diluted, which is qualitatively the same as the situation in the earlier work~\cite{Barreiro:2007hb,Alam:2022rtt}. This illustrates the general point emphasized in Sec.~\ref{sec:curvature_assisted_compactification}: unlike curvature, matter excitations do not merely provide friction, but also induce radion-dependent effective terms that can obstruct stabilization itself. As an illustration of this point, we show the time-dependent effective potential $U_{\rm eff}$~\eqref{Ueff} in Fig.~\ref{fig:time dependent potential}. In this example, the thermal potential appears dominant, and one might naively expect the tracker solution to be difficult to realize. Nevertheless, the radion seems properly stabilized. This is understood as follows: As the curvature dominates the universe, the Hubble friction dominates radion dynamics, and the radion freezes at some value until the tracker solution is realized around the time $U_{\rm th}<U_{\rm vac}$. Therefore, although the thermal potential prevents the realization of the tracker solution, as long as the radion stays at the value smaller than the critical value $b=b_{\rm bar}$ (see Fig.~\ref{fig:vacuum potential}), the tracker solution can eventually be realized. Indeed, for $r_K=300$, the radion freezes at the value $b>b_{\rm bar}$ and therefore the radion stabilization fails. In this sense, the role of the curvature energy density is not just to realize the tracker solution but also to freeze the radion until the tracker solution is realized. 

\begin{figure}
    \centering
    \includegraphics[width=0.8\linewidth]{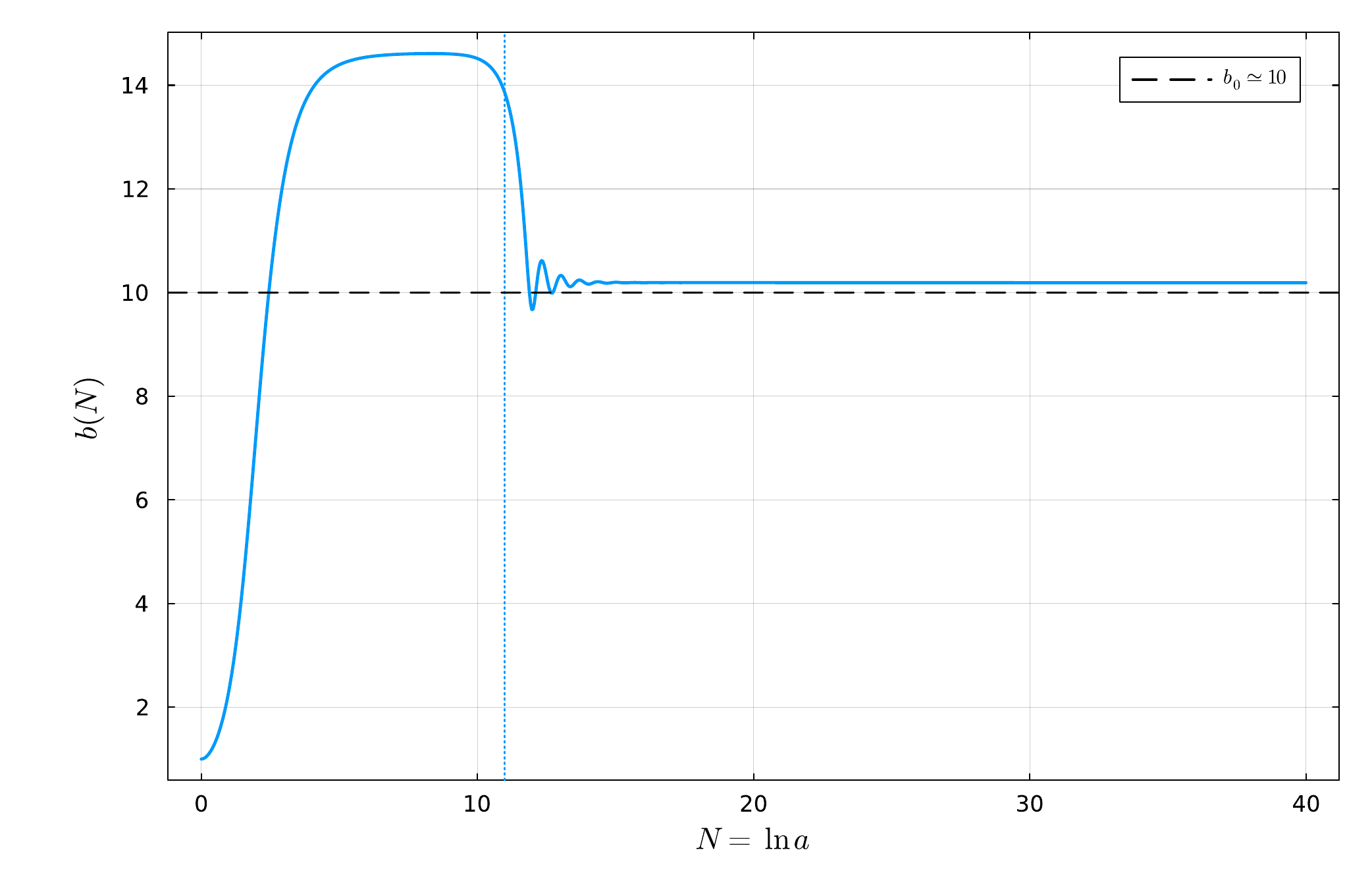}
    \caption{The dynamics of radion in the presence of the inflationary potential. The inflaton potential couples to the radion and shifts the radion local minimum, which is why the radion stays at the value slightly larger than $b_0$. Apart from this shift, the radion dynamics is the same as the tracker solution without inflaton potential.}
    \label{fig:radion trajectory inflation}
\end{figure}

Let us now consider the inflation after radion stabilization. At the time of the radion stabilization, the universe is dominated by the curvature energy density and cannot be consistent with the present universe. The inflaton potential can be thought of as an additional cosmological constant as can be seen from the form of the equation of motion~\eqref{EOMr1}. Accordingly, if we turn on the inflaton potential, there is a possibility to destabilize the radion potential, which is a well-known problem~\cite{Accetta:1986vt,Kallosh:2004yh,Buchmuller:2004tz}. Note that in the presence of the inflaton potential, the radion potential becomes shallower than that in Fig.~\ref{fig:time dependent potential}. This fact should be taken into account when one wants to determine the critical value of the curvature radius below which the moduli stabilization is achieved. As an illustrative example, we have chosen the inflaton potential parameter and the open universe curvature as
\begin{align}
    V_0=2.1\times 10^{-13}, \qquad r_K=250.
\end{align}
Here the height of the inflaton potential $V_0/b_0 \sim 2\times 10^{-14}$ is about $30\%$ of the potential barrier height. As we do not intend to discuss inflationary observables in detail here, we simply fix
$\alpha=1$ in the numerical example, although this value does not reproduce the observed
amplitude of the scalar curvature perturbation. In a single-field slow-roll case after the radion stabilization, one would instead choose $\alpha=O(10^{-4})$ for
$N_* \simeq 50$--$60$, where $N_*$ denotes the number of e-folds before the end of
inflation at which the CMB pivot scale exits the horizon. This change, however, does not
significantly affect the radion dynamics shown below. The shift of the radion minimum
during the inflationary stage is controlled mainly by the plateau height,
$\Delta V \sim V_0/b$, and is therefore essentially independent of $\alpha$ as long as the
inflaton is on the plateau.  We show the radion dynamics in Fig.~\ref{fig:radion trajectory inflation}. The effective cosmological constant shifts the stabilized radion value slightly, but the dynamics is almost the same as the case without inflation. Although the figure shows only the first 40 e-folds, the inflation continues and eventually ends. At the end of inflation, we expect that radion starts to oscillate, and if the radion lifetime is longer than that of inflaton, radion may dominate the universe. We also note that, at the beginning of inflation, the radion significantly oscillates around the local minimum, and such a feature may affect the spectrum of the curvature perturbations. Such a feature resembles the case of the Randall-Sundrum cosmological model discussed in~\cite{Mishra:2025ofh}.

\begin{figure}
    \centering
    \includegraphics[width=1\linewidth]{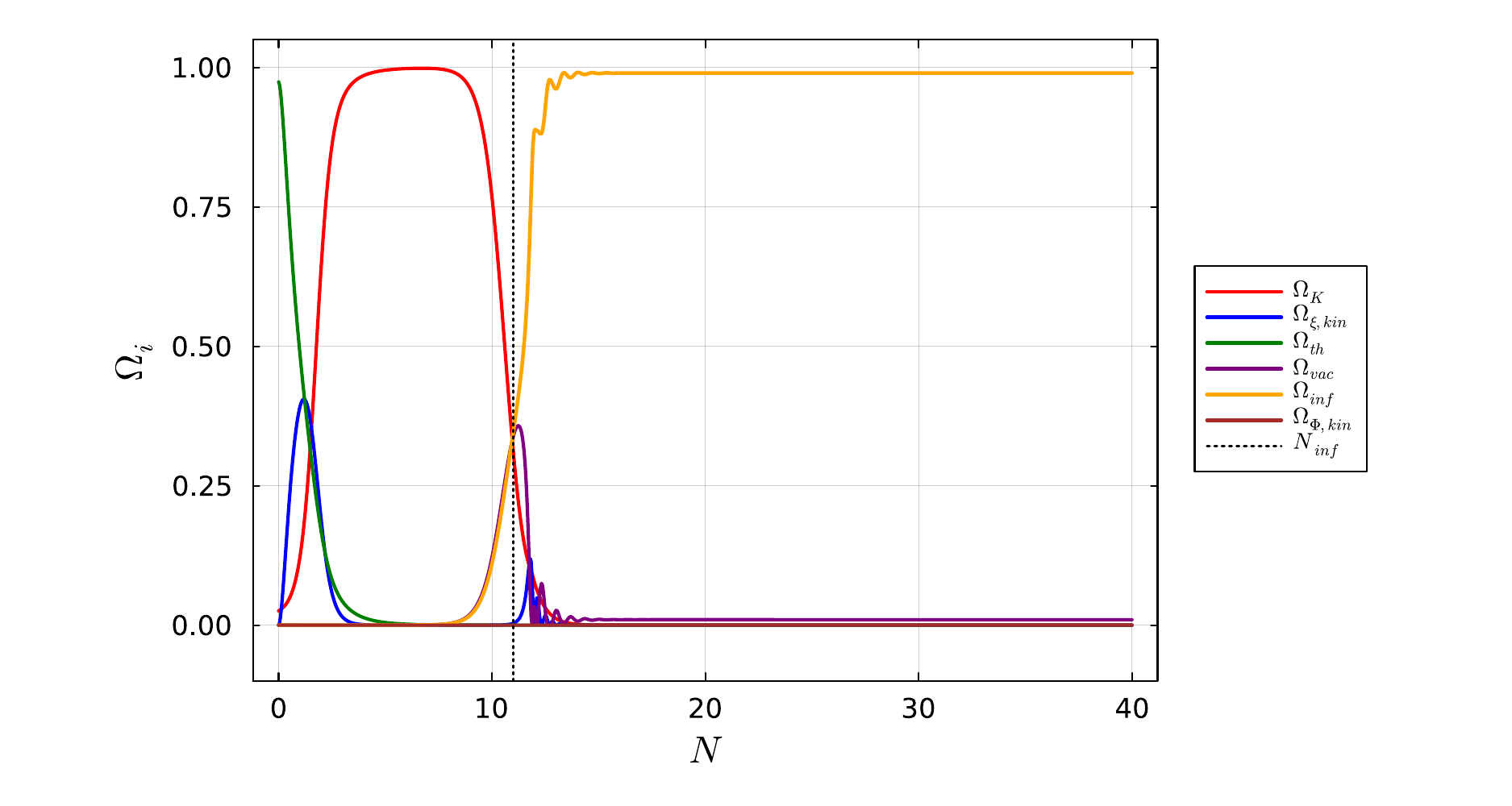}
    \caption{The time dependence of the energy fraction. Each curve corresponds to an energy component divided by the total energy. At the very beginning, the thermal energy dominates the universe but soon the curvature energy dominates. Around the time of the radion stabilization, inflation starts and the curvature energy starts to decay exponentially as expected.}
    \label{fig:energy fraction}
\end{figure}

Let us discuss the time evolution of the energy ratio within our scenario. We define the following quantities,
\begin{align}
   & \Omega_K=\frac{1}{a^2r_K^2H^2}, \qquad \Omega_{\xi,{\rm kin}}=\frac{\dot\xi^2}{6H^2M_{\rm pl}^2},\qquad \Omega_{\rm th}=\frac{U_{\rm th}}{3H^2M_{\rm pl}^2}, \qquad \Omega_{\rm vac}=\frac{U_{\rm eff}-U_{\rm th}}{3H^2M_{\rm pl}^2},\nonumber\\
    &\Omega_{\rm inf}=\frac{V_{\rm inf}}{3H^2M_{\rm pl}^2}, \qquad \Omega_{\Phi,\rm kin}=\frac{\dot\Phi^2}{6H^2M_{\rm pl}^2},
\end{align}
and we show the time dependence of each quantity in Fig.~\ref{fig:energy fraction}. In the first few e-folds, the thermal energy dominates but eventually the curvature energy dominates the universe as expected. Around the time when the radion is trapped, inflaton potential energy also starts to dominate and the curvature energy exponentially decays. The radion energy is subdominant but yet non-vanishing since the radion potential minimum is slightly shifted as mentioned above. Thus, we have shown an explicit numerical example that realizes our scenario.


We finally discuss how long inflation should continue. At the beginning of inflation the energy fraction of the curvature energy is about half. In the present universe, the CMB observation constrains the energy density of the curvature to be $\Omega_{K0}\leq{\cal O}(10^{-3})$~\cite{Planck:2018vyg}. Then, we find the constraints
\begin{align}
    10^{-3}\geq \Omega_{K0}=\Omega_{K}|_{\rm ini}\times e^{-2N_{\rm tot}}\times \left(\frac{a_{\rm end}H_{\rm end}}{a_0 H_0}\right)^{2},
\end{align}
where $\Omega_K|_{\rm ini}$ is the curvature energy ratio at the beginning of inflation, $N_{\rm tot}$ is the total number of e-folds during inflation, and $a_0$, $H_0$ are the Hubble parameter and the scale factor in the present universe while $H_{\rm end}$, $a_{\rm end}$ are evaluated at the end of inflation. Then, we find the constraint on the total number of e-folds as
\begin{align}
    N_{\rm tot}\geq 62.4-\log\left[\left(\frac{\Omega_{K0}}{10^{-3}}\right)^{\frac12}\left(\frac{10^{11}{\rm GeV}}{H_{\rm end}}\right)\left(\frac{T_{\rm reh}}{10^{14}{\rm GeV}}\right)\right]
\end{align}
where we have assumed $\Omega_K|_{\rm ini}=1/2$ and the entropy conservation at the reheating, and taken the effective number of degrees of freedom today as $g_S=3.91$ while at the time of reheating to be $g_S=106.75$. In principle, if inflation continues long enough, our scenario is consistent with the present observational results. However, it also implies that all signatures of higher-dimensional spacetime can be erased by inflation. Since KK modes are too heavy to be excited from vacuum even in the de Sitter background, we do not expect cosmological collider physics type signatures in our model, and possible KK signatures would have to arise from remnants of the initial thermal population. We also note that as we explained above \eqref{rhophi}, gravitational particle production at the moment of radion stabilization does not produce KK particles.\footnote{As a numerical illustration, we note that the Hubble scale at the time of radion stabilization is $H\sim 10^{12}$GeV whereas the KK mass scale is $m_{KK}=\frac{1}{Rb_0^{3/2}}\sim 7.5\times 10^{15}$GeV. Therefore, cosmological KK particle production is negligible. Note that the thermal excitation is introduced as the initial state, and such an excitation is distinguished from the excitation associated with cosmological particle production. Furthermore, in our example, we consider bulk mass, which is slightly smaller than KK mass scale but yet larger than Hubble scale. Therefore, particle production does not occur even for zero modes.} The earlier cosmic expansion as well as that during inflation dilutes the KK particles very rapidly. Therefore, we expect that any signatures of KK particles would be highly suppressed. In this sense, if any signatures of higher-dimensional spacetime may appear within our scenario, they would appear only for the low-$l$ modes in the CMB anisotropy spectrum. We leave observable signatures of this class of models for future work.

\begin{figure}
    \centering
    \includegraphics[width=0.7\linewidth]{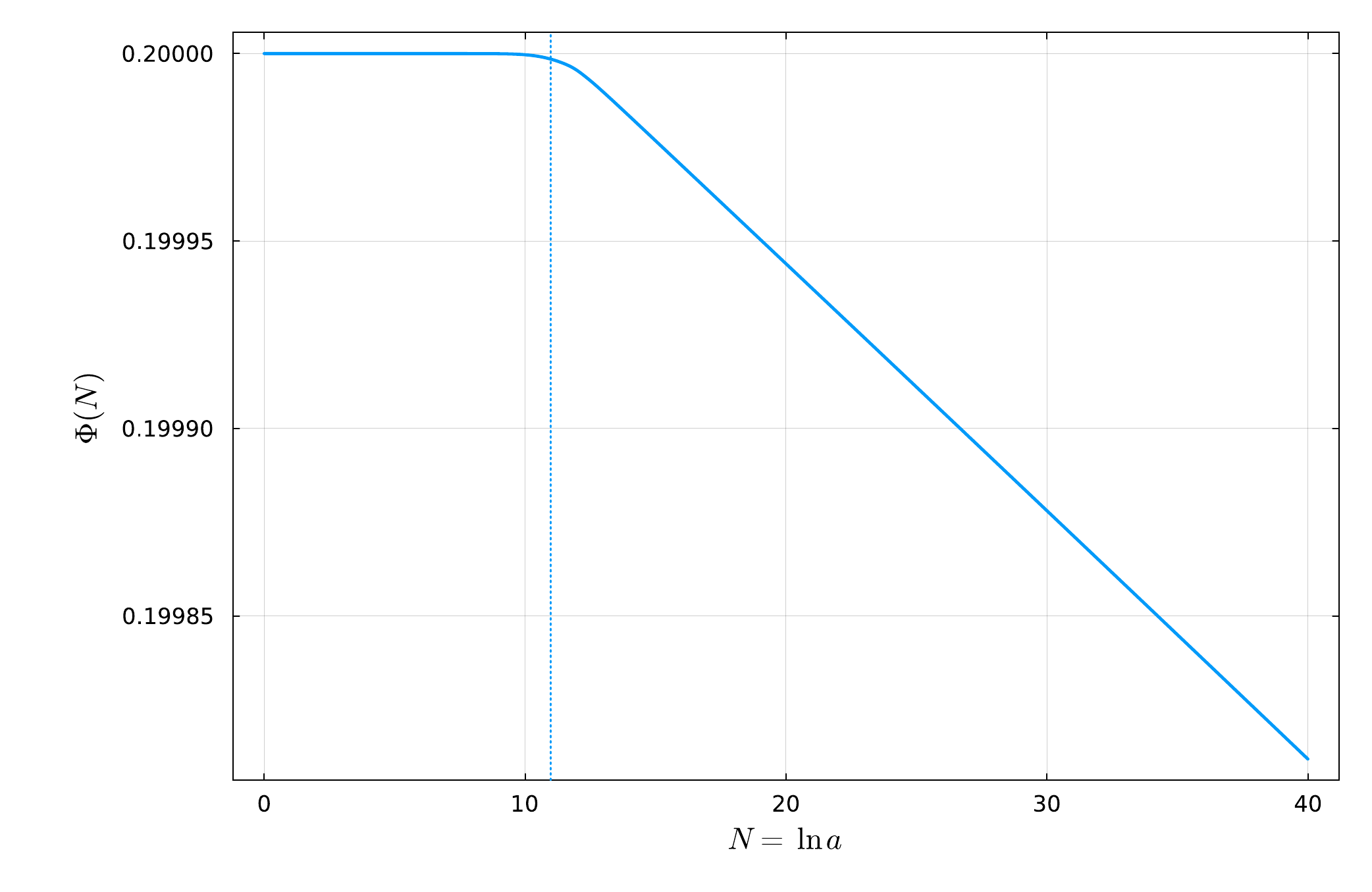}
    \caption{The time evolution of the inflaton field. At the first 10 e-folds, namely during 5D expansion stage, the inflaton freezes on its potential, and only after its potential dominates the universe, the inflaton starts to roll down its potential. }
    \label{fig:inflaton trajectory}
\end{figure}

\subsection{Conditions on inflationary models}

In this section, we briefly summarize the conditions on inflationary models within our setup.
The first condition is that the inflationary energy scale should not destabilize the radion.
Since the inflaton potential appears in the radion equation through $V_{\rm inf}/b$, the
inflationary plateau height must be smaller than the height of the radion potential barrier.
We therefore require
\begin{align}
    \frac{V_{\rm inf}}{b_0}
    \ll
    \Delta V_{\rm barrier}
    \equiv
    V_{\rm eff}(b_{\rm barrier})-V_{\rm eff}(b_0),
    \label{eq:barrier_condition}
\end{align}
where $b_{\rm barrier}$ denotes the position of the local maximum of the radion potential
near the stabilized minimum $b=b_0$. In the numerical example above, the additive constant
is chosen so that $V_{\rm eff}(b_0)$ is much smaller than the barrier height, and hence
$V_{\rm eff}(b_{\rm barrier})$ approximately represents the barrier height in this convention.
This condition has to be maintained during the subsequent inflationary stage. It is also
useful to express the same requirement in terms of the local stability of the radion,
\begin{align}
    H_{\rm inf}^2 \ll m_{\rm rad}^2 ,
\end{align}
where $m_{\rm rad}$ is the canonically normalized radion mass around the stabilized minimum.
In ordinary single-scale potentials, this hierarchy is often correlated with the fact that the
modulus stabilization scale is higher than the inflaton mass scale. However, the precise
relation between the radion and inflaton masses is model-dependent; what is essential for
our scenario is the stability condition above.

Let us next discuss the initial condition for the inflaton. During the early five-dimensional
expanding stage, the energy density is dominated by the thermal and curvature components,
and the inflaton potential is subdominant. In particular, in the parameter regime considered
above, the Hubble scale during the early five-dimensional expansion is much larger than the
Hubble scale during the later inflationary phase after radion stabilization. For a sufficiently
flat plateau potential, the inflaton equation is therefore overdamped. More explicitly, if
\begin{align}
    H^2 M_{\rm pl}^2 \gg \frac{V_{\rm inf}}{b},
    \qquad
    \frac{1}{b}\left|\partial_{\Phi}V_{\rm inf}\right| \ll H^2 M_{\rm pl},
    \qquad
    m_{\Phi,\rm eff}^2 \equiv \frac{1}{b}\partial_{\Phi}^2 V_{\rm inf} \ll H^2 ,
\end{align}
then
\begin{align}
    \ddot{\Phi}+3H\dot{\Phi}+\frac{1}{b}\partial_{\Phi}V_{\rm inf}
    \simeq
    \ddot{\Phi}+3H\dot{\Phi},
    \label{eq:inflaton_freezing}
\end{align}
and the inflaton is effectively frozen by the large Hubble friction. This is precisely what
happens for the plateau-like potential used in our numerical example. The time evolution
of the inflaton field is shown in Fig.~\ref{fig:inflaton trajectory}: during the first ten e-folds, corresponding to the
five-dimensional expansion stage, the inflaton remains almost constant, and it starts to
roll only after its potential energy becomes the dominant component. Thus any initial inflaton velocity is rapidly damped as $\dot\Phi\propto a^{-3}$ in this overdamped regime, and the inflaton approaches an approximately constant value until the potential energy becomes cosmologically relevant.

Therefore, as long as the radion stabilization is insensitive to the inflaton potential in the
sense of Eq.~\eqref{eq:barrier_condition}, the subsequent slow-roll stage is not strongly
constrained by the preceding compactification dynamics. The required choice of the inflaton
initial condition is then comparable to that in the corresponding standard slow-roll model:
the inflaton should initially lie on a sufficiently flat part of the potential, while the large
pre-inflationary Hubble friction keeps it frozen until the compactified four-dimensional
inflationary phase begins.

\section{Conclusions and Discussion}\label{sec:conclusion}
We have proposed a model of higher-dimensional spacetime in which the extra dimensions are not assumed to be already static or stabilized, but are instead required to become stabilized dynamically during cosmological evolution.
This should be contrasted with many four-dimensional effective descriptions, in which stabilized extra dimensions are effectively part of the initial data. Within a controllable semiclassical regime of a five-dimensional toy model, we showed that negative curvature can play a qualitatively distinct role in this problem: because it redshifts more slowly than radiation- or matter-like components, it can sustain tracker-like radion evolution and keep the modulus under sufficient control until the thermal obstruction is diluted and the vacuum minimum becomes effective. It is worth emphasizing again that negative curvature does not induce an effective moduli potential at leading order, whereas radiation and matter generally do, particularly when the moduli originate from gravitational degrees of freedom. In this sense, the present example provides a concrete proof of principle that a genuinely higher-dimensional expanding phase can be connected dynamically to a later effectively four-dimensional inflationary universe.

Although our analysis has been carried out in a simple 5D toy model, we view it primarily as a calculable realization of a broader mechanism rather than as a claim about five dimensions in particular. It would be interesting to investigate whether similar curvature-assisted pre-inflationary compactification can be realized in more general higher-dimensional settings, especially in models motivated by string compactifications. In such extensions, however, additional ingredients---including the structure of the internal space, anisotropies, backreaction, and the treatment of quantum matter in more complicated time-dependent backgrounds---may become important, so the generalization is not entirely automatic and deserves separate study.

More broadly, the present work proposes a framework in which extra dimensions are not assumed to be already static or stabilized as part of the initial data of a four-dimensional effective theory, but are instead required to become stabilized dynamically during cosmological evolution. Our analysis also illustrates how difficult it can be to obtain the present effectively four-dimensional universe from a higher-dimensional Big-Bang phase. In the explicit example studied here, successful radion stabilization requires a sufficiently large initial open curvature, which highlights the important role of the non-compact spatial geometry. In this sense, the mechanism realized in the present setup is specific to an open-universe background rather than being generic for arbitrary spatial topology.

In the present toy model, any direct signature of extra dimensions is expected to be most relevant near the onset of the final inflationary stage. Since the subsequent inflation dilutes both the open curvature and the thermally populated KK modes, observable imprints are likely to be confined, if present at all, to the largest scales. A more detailed analysis of possible signatures---for example low-$k$ features, transition-induced oscillations, or correlated signals associated with remnant KK excitations---is left for future work.

\subsection{Speculative Comments}

Finally, let us comment on a speculative extension of the present mechanism. What is essential for the tracker mechanism in our scenario is the presence of a sufficiently slowly redshifting component, exemplified here by the open-universe curvature with $\rho \propto a^{-2}$. More generally, one may ask whether other ingredients with similar scaling behavior, such as certain extended objects or wrapped branes, could play an analogous role in a higher-dimensional pre-inflationary phase. Indeed, in the context of string/brane gas cosmology, it is known that extended objects indeed lead to rather nontrivial forces on the moduli fields (see \cite{Battefeld:2005av,Brandenberger:2011et} and references therein). If a later inflationary stage follows, remnants of such components may be sufficiently diluted to remain compatible with present observations. Whether this can be realized in a controlled way, however, is strongly model dependent and may be sensitive to issues such as anisotropic stress, wrapping configurations, and backreaction. We therefore leave this question as an interesting direction for future work.

\section*{Acknowledgments}
Y.Y. is supported by IBS under the project code, IBS-R018-Y3-2026-a00. 
The author used ChatGPT as an auxiliary tool for improving the wording of parts of the manuscript, preparing Fig.~\ref{fig:scenario_concept}, and developing numerical code. All outputs were checked and edited by the author, who takes full responsibility for the scientific content, calculations, interpretations, and final manuscript.

\appendix 
\section{Quantum Fields in 5D Open FRW Background}\label{quantum field in openFRW}
We summarize the properties of the mode functions in the 5D open FRW background~\eqref{5D metric}. 
\subsection{Scalar Fields}
For the open universe, it is useful to take
\begin{align}
    r=r_K \sinh \chi,
\end{align}
namely the coordinate is given by
\begin{align}
    ds^2_{\rm op}=\frac{1}{b(t)}\left(-dt^2+a^2(t)r_K^2\left(d\chi^2+\sinh^2\chi(d\theta^2+\sin^2\theta d\varphi^2)\right)\right)+b^2(t)dy^2.
\end{align}
A real scalar field action is now 
\begin{align}
    S=\frac12\int d^4x dy\sqrt{-g_4}\left[\dot\phi^2-\frac{1}{a^2r_K^2}\left\{(\partial_\chi\phi)^2+\frac{1}{\sinh^2\chi}\left((\partial_\theta\phi)^2+\frac{1}{\sin^2\theta}(\partial_\varphi\phi)^2\right)\right\}-\frac{1}{b^3}(\partial_y\phi)^2-\frac{m^2}{b}\phi^2\right]
\end{align}
where $\sqrt{-g_4}=a^3 r_K^3\sinh^2\chi\sin\theta$ is the volume element of the 4D FRW spacetime. We KK-expand the scalar field as\footnote{We introduce the scale factor in the KK expansion, which removes overall scale factors in the action.}
\begin{align}
    \phi(x,y)=\frac{1}{a^{\frac32}\sqrt{2\pi R}}\sum_{n\in \mathbb Z}\phi_n(x)e^{\ri \frac{ny}{R}}
\end{align}
and perform $y$-integration, which yields (after integration by part)
\begin{align}
    S=\frac12\sum_n\int d^4x\sqrt{g_3}\left[\left|\dot\phi_n\right|^2-\frac{1}{a^2r_K^2}\left\{\left|\partial_\chi\phi_n\right|^2+\frac{1}{\sinh^2\chi}\left(|\partial_\theta\phi_n|^2+\frac{1}{\sin^2\theta}|\partial_\varphi\phi_n|^2\right)\right\}-m_n^2(t)|\phi_n|^2\right],
\end{align}
where $m_n^2(t)\equiv \frac{m^2}{b}+\frac{n^2}{R^2b^3}-\frac94H^2-\frac32\dot{H}$ and $\sqrt{g_3}=r_K^3\sinh^2\chi\sin\theta$ is the volume element of the static open universe. We note that $\phi_n=\phi_{-n}^*$ following the reality of $\phi$ and the sum is taken over all integers.\footnote{For $n\neq 0$, the sum double-counts each KK mode of the level $|n|$, which removes the overall factor of $1/2$.} We further expand $\phi_n$ as
\begin{align}
    \phi_n(x)=\int_0^\infty dq\sum_{l=0}^\infty\sum_{m=-l}^l\phi_{nqlm}(t)P_{q,l}(\chi)Y_{l,m}(\theta,\varphi)
\end{align}
where $Y_{lm}(\theta,\varphi)$ is the spherical harmonics, satisfying
\begin{align}
    \frac{1}{\sin\theta}\partial_\theta\left(\sin\theta\partial_\theta Y_{l,m}\right)+\frac{1}{\sin^2\theta}\partial_\varphi^2Y_{l,m}=-l(l+1)Y_{l,m},
\end{align}
and note that $Y_{l,m}^*(\theta,\varphi)=(-1)^m Y_{l,-m}(\theta,\varphi)$.
The radial mode function $P_{q,l}(\chi)$, which is real, satisfies
\begin{align}
   - \frac{1}{\sinh^2\chi}\partial_\chi(\sinh^2\chi \partial_\chi P_{q,l})+\frac{l(l+1)}{\sinh^2\chi}P_{q,l}=(q^2+1) P_{q,l}
\end{align}
where $q\geq0$ is the continuous eigenvalue, which can be confirmed as follows. Introducing $u_{ql}(\chi)=\sinh\chi P_{q,l}(\chi)$, the radial equation can be rewritten as $-u''+\frac{l(l+1)}{\sinh^2\chi}u=q^2u$. This is the Schr\"odinger equation with a non-negative potential, and therefore, its energy spectrum should be positive, namely $q^2\geq0$. Now, we have formally written the eigenvalue as $q^2$, and therefore the value $q,-q$ should not be distinguished as an independent eigenvalue. Therefore, without loss of generality, we may take $q\geq0$. The explicit form of the solution is
\begin{align}
    P_{q,l}(\chi)=N_{q,l}(\sinh\chi)^l{}_2F_1\left(l+1+\ri q,l+1-\ri q; l+\frac32;-\sinh^2\frac{\chi}{2}\right),
\end{align}
where $N_{q,l}$ is a normalization factor. Nevertheless, the explicit forms of each function are not necessary in most computations but it is important to note the normalization conditions
\begin{align}
&\int_{S^2} d\Omega Y_{l,m}Y^*_{l',m'}= \int_{0}^\pi d\theta\int_0^{2\pi} d\varphi \sin\theta Y_{l,m}Y^*_{l',m'}=\delta_{l,l'}\delta_{m,m'},\\
&\int_0^{\infty} d\chi r_K^3\sinh^2\chi P_{q,l}P_{q',l}=\delta(q-q').
\end{align}
For notational simplicity, we define
\begin{align}
    Q_{qlm}(\chi,\theta,\varphi)\equiv P_{q,l}(\chi)Y_{l,m}(\theta,\varphi).\label{defQ}
\end{align}
The completeness relation is
\begin{align}
    \int_0^\infty d q\sum_{l=0}^\infty \sum_{m=-l}^l  Q_{qlm}(\chi,\theta,\varphi)Q_{qlm}^*(\chi',\theta',\varphi')=\frac{1}{\sqrt{g_3}}\delta(\chi-\chi')\delta(\theta-\theta')\delta(\varphi-\varphi').
\end{align}
With these base functions, the quantum field operator $\hat{\phi}_n(x)$ can be expanded as
\begin{align}
    \hat{\phi}_{n}(x)=\int_0^\infty dq\sum_{l=0}^\infty\sum_{m=-l}^l\left[\hat{a}_{nqlm} f_{n,q}(t)Q_{qlm}(\chi,\theta,\varphi)+\hat{a}^\dagger_{-nqlm}f^*_{n,q}(t)Q^*_{qlm}(\chi,\theta,\varphi)\right],
\end{align}
where $\hat{a}_{nqlm},\hat{a}_{nqlm}^\dagger$ are annihilation, creation operators satisfying
\begin{align}
    \left[\hat{a}_{nqlm},\hat{a}_{n'q'l'm'}^\dagger\right]=\delta_{n,n'}\delta(q-q')\delta_{l,l'}\delta_{m,m'}.
\end{align}
The commutation relation of the creation, annihilation operators are consistent with the canonical commutation relation
\begin{align}
    \left[\hat{\phi}_n(t,\chi,\theta,\varphi),\hat{\Pi}_{n'}(t,\chi',\theta',\varphi')\right]=\frac{\ri\delta_{nn'}\delta(\chi-\chi')\delta(\theta-\theta')\delta(\varphi-\varphi')}{\sqrt{g_3}},
\end{align}
where $\hat{\Pi}_n\equiv \dot{\hat\phi}_n^\dagger=\dot{\hat\phi}_{-n}$. Here, we have imposed the normalization for the time mode function as
\begin{align}
    f_{n,q}(t)\dot{f}^*_{n,q}(t)-f^*_{n,q}(t)\dot{f}_{n,q}(t)=\ri,
\end{align}
which holds at any time $t$ as long as the mode function is the solution of the mode equation
\begin{align}
    \ddot{f}_{n,q}(t)+\omega_{n,q}^2(t)f_{n,q}(t)=0,
\end{align}
where the time-dependent one-particle energy is defined as
\begin{align}
    \omega^2_{n,q}(t)\equiv \frac{q^2+1}{a^2(t)r_K^2}+m_n^2(t).
\end{align}

In the practical computations, we encounter the spectral projector on $H^3$ defined as
\begin{align}
    \Pi_{q}(\bm x,\bm x')\equiv \sum_{l=0}^\infty\sum_{m=-l}^lQ_{qlm}(\chi,\theta,\varphi)Q^*_{qlm}(\chi',\theta',\varphi'),
\end{align}
where $\bm x=(\chi,\theta,\varphi)^t$. From the completeness relation, we find $\int dq \Pi_q(\bm x,\bm x')=\delta(\bm x-\bm x')/\sqrt{g_3}$. Due to the spatial translational symmetry, the spectral projection depends only on the distance of the two points $\bm x,\bm x'$. In particular, we are interested in the coincidence limit $\bm x\to\bm x'$, which we call the spectral density
\begin{align}
    \rho_{H^3}(q)\equiv \lim_{\bm x'\to \bm x}\Pi_{q}(\bm x,\bm x')=\lim_{\bm x'\to \bm 0 }\Pi_{q}(\bm 0,\bm x')\label{scalar spectral}
\end{align}
Therefore, we consider the limit that both of them are close to the origin $\chi,\chi'\sim 0$. For $\chi\to 0$, only $l=0$ mode dominates in the sum over $l$, and therefore, we focus on $l=0$. The radial mode function for $l=0$ satisfies $u''_{q0}(\chi)+q^2u_{q0}(\chi)=0$, which implies $P_{q,0}(\chi)=N_{q,0}\frac{\sin(q\chi)}{\sinh\chi}$. The normalization condition reads $N_{q,0}N_{q',0} \int_0^\infty d\chi r_K^3 \sin(q\chi)\sin(q'\chi)=\delta(q-q')$, which leads to $N_{q,0}=\sqrt{\frac{2}{\pi r_K^3}}$. Now, the spectral density becomes
\begin{align}
    \rho_{H^3}(q)=\lim_{\chi,\theta,\varphi\to 0}N_{q,0}^2\left(\frac{\sin(q\chi)}{\sinh\chi}\right)^2|Y_{0,0}(\theta,\varphi)|^2=\frac{q^2}{2\pi^2r_K^3}.\label{scalar spectral2}
\end{align}

\subsection{Spinor Fields}
Next, we discuss quantization of spinor fields. We use the convention in \cite{Wess:1992cp} where the 4D Pauli matrices are given by
\begin{align}
    \sigma^0=\left(\begin{array}{cc}
    -1&0\\
    0&-1\end{array}\right),\quad \sigma^1=\left(\begin{array}{cc}
    0&1\\
    1&0\end{array}\right),\quad \sigma^2=\left(\begin{array}{cc}
    0&-\ri\\
    \ri&0\end{array}\right),\quad \sigma^3=\left(\begin{array}{cc}
    1&0\\
    0&-1\end{array}\right),
\end{align}
$\bar{\sigma}^0=\sigma^0$ and $\sigma^i=-\bar{\sigma}^i$ ($i=1,2,3$). A 5D spinor (=4D Dirac spinor) is represented as
\begin{align}
    \Psi=\left(\begin{array}{c}\psi_\alpha\\ \zeta^{\dagger\dot\alpha}\end{array}\right),
\end{align}
where $\psi_\alpha$ and ${\zeta}^{\dagger\dot\alpha}$ are two-component spinors. We define the $\gamma$-matrices as
\begin{align}
    \gamma^\mu=\ri\left(\begin{array}{cc}
      0   &  \sigma^\mu\\
       \bar{\sigma}^\mu  &0 
    \end{array}\right),
\end{align}
by the 4D Pauli matrices $\{\sigma^\mu\}$ defined above. We also define $\gamma^5$ as
\begin{align}
    \gamma^5=\ri\gamma^0\gamma^1\gamma^2\gamma^3=\left(\begin{array}{cc}
       {\bm 1}_{2\times2}  &0  \\
        0 &  -{\bm 1}_{2\times2}
    \end{array}\right).
\end{align}
The vielbein satisfying $e_M^a\eta_{ab}e^{b}_N=g_{MN}$ are explicitly given by (in the 1-form notation)
\begin{align}
    e^{0}=\frac{1}{\sqrt{b}}dt,\quad e^1=\frac{ar_K}{\sqrt{b}}d\chi,\quad e^2=\frac{ar_K\sinh\chi}{\sqrt b}d\theta,\quad e^3=\frac{ar_K\sinh\chi\sin\theta}{\sqrt{b}}d\varphi,\quad  e^5=b dy,
\end{align}
and non-vanishing spin connection 1-forms are
\begin{align}
    &\omega^{i0}=-\sqrt{b}\left(H-\frac12H_b\right)e^i \ (\text{for }i=1,2,3), \quad \omega^{i1}=\frac{\sqrt b \coth\chi}{ar_K}e^i\ (\text{for }i=2,3), \nonumber\\
    &\omega^{32}=\frac{\sqrt b \cot\theta}{ar_K\sinh\chi}e^3, \quad \omega^{50}=-\sqrt{b}H_b e^5
\end{align}
Now a 5D spinor action is written as
\begin{align}
    S=&-\int d^4x\int_{0}^{2\pi R}dy\sqrt{-g}\left[\bar{\Psi}\Gamma^a e_a^M D_M\Psi-m\bar\Psi\Psi\right]\nonumber\\
    =&-\int d^4x\int_{0}^{2\pi R}dy\sqrt{g_3}\Biggl[\bar\Psi\Biggl\{\gamma^0\partial_t+\frac{1}{ar_K}\gamma^1\left(\partial_\chi+\coth\chi\right)+\frac{1}{ar_K\sinh\chi}\gamma^2\left(\partial_\theta+\frac12\cot\theta\right)\nonumber\\
    &\hspace{3.5cm}+\frac{1}{ar_K\sinh\chi\sin\theta}\gamma^3\partial_\varphi+\frac{1}{b^{\frac32}}\gamma^5\partial_y\Biggr\}\Psi-\frac{m}{\sqrt{b}}\bar{\Psi}\Psi\Biggr],
\end{align}
where we have rescaled the Dirac spinor as $\Psi\to a^{-\frac32}b^{\frac14}\Psi$ in the second equality. In terms of the two-component spinors, the action becomes
\begin{align}
    S=&\sum_{n\in\mathbb Z}\int d^4x\Biggl[-\ri\psi_n^\dagger\bar\sigma^0\dot\psi_n-\ri\zeta_n\sigma^0\dot{\zeta}^\dagger_n-\frac{\ri}{ar_K}\left\{\psi_n^\dagger\bar\sigma^1D_\chi\psi_n+\zeta_n\sigma^1D_\chi\zeta_n^\dagger\right\}\nonumber\\
    &\qquad-\frac{\ri}{ar_K\sinh\chi}\left\{\psi_n^\dagger\bar\sigma^2D_\theta\psi_n+\zeta_n\sigma^2D_\theta\bar{\zeta}_n\right\}-\frac{\ri}{ar_K\sinh\chi\sin\theta}\left\{\psi^\dagger_n\bar\sigma^3\partial_\varphi\psi_n+\zeta_n\sigma^3\partial_\varphi\zeta^\dagger_n\right\}\nonumber\\
    &\qquad-M_n\zeta_n\psi_n-\bar{M}_n\psi^\dagger_n\zeta_n^\dagger\Biggr]\nonumber\\
    =&\sum_{n\in\mathbb Z}\int d^4x\Biggl[-\ri\psi_n^\dagger\bar\sigma^0\dot\psi_n-\ri\zeta_n\sigma^0\dot{\zeta}^\dagger_n-\frac{\ri}{ar_K}\left\{\psi_n^\dagger\slashed{
    D}_{H^3}\psi_n+\zeta^\dagger_n\slashed{D}_{H^3}\zeta_n\right\}-M_n\zeta_n\psi_n-\bar{M}_n\zeta^\dagger_n\psi_n^\dagger\Biggr]\label{2spinordecomp}
\end{align}
where $D_\chi=\partial_\chi+\coth\chi$, $D_\theta=\partial_\theta+\frac12\cot\theta$, $M_n=\frac{m}{\sqrt b}-\frac{\ri n}{b^{\frac32}R}$, and we have introduced the Dirac operator on hyperbolic 3-space $H^3$,
\begin{align}
    \slashed{D}_{H^3}\psi\equiv \bar\sigma^1D_\chi\psi+\frac{1}{\sinh\chi}\left\{\bar\sigma^2 D_\theta\psi+\frac{1}{\sin\theta}\bar\sigma^3\partial_\varphi\psi\right\}.\label{H3Dirac}
\end{align}
The KK expansion of the spinor field is defined as
\begin{align}
    \Psi=\sum_{n\in \mathbb Z}\left(\begin{array}{c}\psi_n\\ \zeta_n^\dagger\end{array}\right)e^{\ri\frac{n y}{R}}.
\end{align}

We need several steps to derive the eigenspinor for the Dirac operator on $H^3$. See~\cite{Camporesi:1995fb} for more general discussions. First, we consider the $S^2$ eigenspinor. We consider a particle with the orbital angular momentum $|l,m_l\rangle$ and the spin angular momentum $\left|\frac12,\pm\frac12\right\rangle$. The total angular momentum state $|j,m\rangle$ can be written as
\begin{align}
    |j,m\rangle=\left|l,m-\frac12\right\rangle\left|\frac12,\frac12\right\rangle\left\langle l,m;\frac12,\frac12\right|j,m\Biggr\rangle+\left|l,m+\frac12\right\rangle\left|\frac12,-\frac12\right\rangle \left\langle l,m;\frac12,-\frac12\right|j,m\Biggr\rangle.
\end{align}
The total angular momentum $j$ can take $j=l\pm 1/2$, and the Clebsch-Gordan coefficients yield the concrete expression,
\begin{align}
   & \left|l+\frac12,m\right\rangle=\sqrt{\frac{l+m+\frac12}{2l+1}}\left|l,m-\frac12\right\rangle\left|\frac12,\frac12\right\rangle+\sqrt{\frac{l-m+\frac12}{2l+1}}\left|l,m+\frac12\right\rangle\left|\frac12,-\frac12\right\rangle,\\
    &\left|l-\frac12,m\right\rangle=-\sqrt{\frac{l-m+\frac12}{2l+1}}\left|l,m-\frac12\right\rangle\left|\frac12,\frac12\right\rangle+\sqrt{\frac{l+m+\frac12}{2l+1}}\left|l,m+\frac12\right\rangle\left|\frac12,-\frac12\right\rangle.
\end{align}
Therefore, the wave function of $|l\pm1/2,m\rangle$ are given by (with $\left|\frac12,\frac12\right\rangle =(1,0)^t$, $\left|\frac12,-\frac12\right\rangle =(0,1)^t$ where ${}^t$ denotes transpose)
\begin{align}
    \xi^+_{j,m}(\theta,\varphi)=&\left(\begin{array}{c}\sqrt{\frac{j+m}{2j}}Y_{j-\frac12,m-\frac12}(\theta,\varphi)\\ \sqrt{\frac{j-m}{2j}}Y_{j-\frac12,m+\frac12}(\theta,\varphi)\end{array}\right),\\
    \xi^-_{j,m}(\theta,\varphi)=&\left(\begin{array}{c}-\sqrt{\frac{j+m+1}{2j+2}}Y_{j+\frac12,m-\frac12}(\theta,\varphi)\\ \sqrt{\frac{j-m+1}{2j+2}}Y_{j+\frac12,m+\frac12}(\theta,\varphi)\end{array}\right).
\end{align}
Notice that the state $|j,m\rangle$ are eigenvector of the operator $\hat{D}=(2\hat{\bm s}\cdot\hat{\bm L}+1)$ where $\hat{\bm s}=\frac12\bm \sigma$ is the spin angular momentum operator ${\bm L}=-\ri\hat{\bm r}\times \hat{\bm \nabla}$ is the orbital angular momentum. Indeed, we find $\hat{D}=\hat{\bm J}^2-\hat{\bm S}^2-\hat{\bm L}^2+1$, and the wave functions are the eigenfunction of $\hat{D}$:
\begin{align}
    \hat{D}\xi_{j,m}^\pm=\pm\left(j+\frac12\right)\xi_{j,m}^\pm.
\end{align}
Noting that ${\bm \nabla}={\bm n}_r\partial_r+{\bm n}_\theta\frac{1}{r}\partial_\theta+{\bm n}_\varphi\frac{1}{r\sin\theta}\partial_\varphi$ with the unit vector ${\bm n}_i$, we find
\begin{align}
    \hat{D}=-\ri\sigma_\varphi\partial_\theta+\ri\sigma_\theta\frac{1}{\sin\theta}\partial_{\varphi}+1,
\end{align}
where $\sigma_{\varphi}={\bm n}_{\varphi}\cdot{\bm \sigma}$ and $\sigma_{\theta}={\bm n}_{\theta}\cdot{\bm \sigma}$.
Now we consider the rotation of the spin eigenvector by $\hat{U}=e^{-\frac{\ri}{2}\varphi\sigma_3}e^{-\frac{\ri}{2}\theta\sigma_2}$, which rotates the Pauli matrix as $\hat{U}\sigma_3\hat{U}^\dagger={\bm n}_r\cdot{\bm \sigma}$, $\hat{U}\sigma_1\hat{U}^\dagger=\sigma_\theta$, $\hat{U}\sigma_2\hat{U}^\dagger=\sigma_\varphi$. We also introduce a constant rotation matrix $\hat{V}=e^{\frac{\ri\pi}{4}\sigma_2}$. With the rotation matrices, we define
\begin{align}
    \Xi^{\pm}_{j,m}(\theta,\varphi)=\hat{V}^\dagger\hat{U}^\dagger\xi^{\pm}_{j,m}(\theta,\varphi).
\end{align}
The rotated wave functions $\Xi^\mp_{j,m}$ are the eigenspinor of $\sigma_r={\bm n}_r\cdot{\bm \sigma}$ and the rotated operator $\hat{\cal D}=-\ri\hat{V}^\dagger\hat{U}^\dagger\hat{D}\hat{U}\hat{V}$, which is given by
\begin{align}
    \hat{\cal D}=\bar{\sigma}_2D_\theta+\bar{\sigma}_3\frac{1}{\sin\theta}\partial_\varphi,
\end{align}
which is the Dirac operator on $S^2$ shown in \eqref{H3Dirac}. 

Let us construct the eigenspinor on $H^3$. Notice that $\{\sigma_1,\hat{\cal D}\}=0$, which leads to the fact that $\sigma_1\Xi^{\pm}_{j,m}\propto \Xi^{\mp}_{j,m}$ since 
\begin{align}
    \hat{\cal D}(\sigma_1\Xi^\pm_{j,m})=-\sigma_1\hat{\cal D}\Xi^{\pm}_{j,m}=\pm\ri\left(j+\frac12\right)\sigma_1\Xi^{\pm}_{j,m}
\end{align}
while $\hat{\cal D}\Xi^{\pm}_{j,m}=\mp\ri\left(j+\frac12\right)\Xi^{\pm}_{j,m}$. Therefore, in the following, we take \begin{align}
    \bar\sigma_1\Xi^{\pm}_{j,m}=\Xi^{\mp}_{j,m},
\end{align}
which is achieved by choosing appropriate phases. Since the Dirac operator~\eqref{H3Dirac} contains $\sigma_1$, we expect that the eigenspinor should consist of both $\Xi^{\pm}_{j,m}$, which implies the ansatz,
\begin{align}
    \psi_{j,m}=F_j(\chi)\Xi_{j,m}^+(\theta,\varphi)+G_j(\chi)\Xi_{j,m}^-(\theta,\varphi). 
\end{align}
Assuming the eigenequation $\ri\slashed{D}_{H^3}\psi_{j,m}=\lambda\psi_{j,m}$, we obtain
\begin{align}
    \left\{\begin{array}{c}D_\chi G_j-\frac{\ri\left(j+\frac12\right)}{\sinh\chi}F_j=-\ri\lambda F_j\\ D_\chi F_j+\frac{\ri \left(j+\frac12\right)}{\sinh\chi}G_j=-\ri\lambda G_j\end{array}\right.
\end{align}
Note that, since the Dirac operator $\ri\slashed{D}_{H^3}$ is Hermitian, $\lambda$ is a real eigenvalue. Introducing $G=g(\chi)/\sinh\chi$, $F=f(\chi)/\sinh\chi$, $f=(p+q)/2$, $g=\ri(p-q)/2$, we can rewrite the above equation as
\begin{align}
  \left\{\begin{array}{c} p'-\frac{\kappa}{\sinh\chi}p=-\lambda q\\ q'+\frac{\kappa}{\sinh\chi}q=\lambda p\end{array}\right. 
\end{align}
where $\kappa=j+\frac12$. Furthermore, we find that $p=(\tanh \chi/2)^\kappa P(\chi)$, $q=(\tanh\chi/2)^{-\kappa}Q(\chi)$ satisfy 
\begin{align}
    \left\{\begin{array}{c}P'=-\lambda \left(\tanh\frac{\chi}{2}\right)^{-2\kappa}Q\\ Q'=\lambda \left(\tanh\frac{\chi}{2}\right)^{+2\kappa}P\end{array}\right.
\end{align}
which is reduced as a single 2nd order differential equation (after some computation)
\begin{align}
    z(z-1)\frac{d^2}{dz^2}P+\left(z-\frac12-\kappa\right)\frac{d}{dz}P+\lambda^2P=0,
\end{align}
where we have introduced a new variable $z\equiv -\sinh^2\frac{\chi}{2}$. The solution regular at $\chi=0$ is given by
\begin{align}
    P(z)={}_2F_1\left(\ri\lambda,-\ri\lambda;\kappa+\frac12,z\right),
\end{align}
and $Q(\chi)=-\lambda^{-1}\left(\tanh\chi/2\right)^{2\kappa}\partial_\chi P$. Following the definitions of each quantity, we finally obtain
\begin{align}
    F_{\lambda,j}(\chi)=&\frac{N_{\lambda,j}}{2}\left(\tanh\frac{\chi}{2}\right)^{j+\frac12}\left[\frac{1}{\sinh\chi}{}_2F_1(\ri\lambda,-\ri\lambda;j+1;z)+\frac{\lambda}{2(j+1)}{}_2F_1(1+\ri\lambda,1-\ri\lambda;j+2;z)\right]\\
    G_{\lambda,j}(\chi)=&\frac{\ri N_{\lambda,j}}{2}\left(\tanh\frac{\chi}{2}\right)^{j+\frac12}\left[\frac{1}{\sinh\chi}{}_2F_1(\ri\lambda,-\ri\lambda;j+1;z)-\frac{\lambda}{2(j+1)}{}_2F_1(1+\ri\lambda,1-\ri\lambda;j+2;z)\right]
\end{align}
where $N_{\lambda,j}$ is a normalization constant. The normalization conditions of the basis functions are as follows. The $S^2$ spinors satisfy
\begin{align}
    \int_{S^2} d\Omega \Xi^{s\dagger}_{j,m}(\theta,\varphi)\Xi^{s'}_{j',m'}(\theta,\varphi)=\delta_{ss'}\delta_{jj'}\delta_{mm'},
\end{align}
where $s,s'=+,-$, $j,j'=\frac12,\frac32,\cdots$, $m=-j,-j+1,\cdots, j-1,j$. The completeness relation reads
\begin{align}
    \sum_{s,j,m}\Xi^s_{j,m}(\theta,\varphi)\Xi_{j,m}^{s\dagger}(\theta',\varphi')=\frac{\delta(\theta-\theta')\delta(\varphi-\varphi')}{\sin\theta}.
\end{align}
Now, we define the $H^3$ eigenspinor 
\begin{align}
    \Theta^S_{\lambda,j,m}(\chi,\theta,\varphi)=F^S_{\lambda,j}(\chi)\Xi^+_{j,m}(\theta,\varphi)+G^S_{\lambda,j}(\chi)\Xi^-_{j,m}(\theta,\varphi),\label{defTheta}
\end{align}
where we have restricted $\lambda\geq0$ but instead we have introduced a label $S=\pm$, which means
\begin{align}
    \ri\slashed{D}_{H^3}\Theta^S_{\lambda,j,m}=S\lambda \Theta^S_{\lambda,j,m}.
\end{align}In this case, the normalization condition can be written as
\begin{align}
    \int d\chi d\theta d\varphi\sqrt{g_3}\Theta^{S\dagger}_{\lambda,j,m}(\chi,\theta,\varphi)\Theta^{S'}_{\lambda',j',m'}(\chi,\theta,\varphi)=\delta(\lambda-\lambda')\delta_{SS'}\delta_{jj'}\delta_{mm'},
\end{align}
and the completeness relation is
\begin{align}
    \int_0^\infty d\lambda \sum_{S,j,m}\Theta_{\lambda,j,m}^S(\chi,\theta,\varphi)\Theta^{S\dagger}_{\lambda,j,m}(\chi',\theta',\varphi')=\frac{\delta(\chi-\chi')\delta(\theta-\theta')\delta(\varphi-\varphi')}{\sqrt{g_3}}.
\end{align}
Finally, as in the scalar field case, we need the spectral density on $H^3$ for spinor fields. The spectral projector of the spinor field is defined as
\begin{align}
    \Pi^f_{\lambda}(\bm x,\bm x')\equiv \sum_{S=\pm}\sum_{j,m}\Theta^S_{\lambda,j,m}(\chi,\theta,\varphi)\Theta^{S\dagger}_{\lambda,j,m}(\chi',\theta',\varphi').
\end{align}
The spectral density is defined as the trace of the spectral projector
\begin{align}
    \rho^f_{H^3}(\lambda)=\lim_{\bm x'\to \bm x}{\rm tr}\left(\Pi^f_{\lambda}(\bm x,\bm x')\right)=\lim_{\bm x'\to \bm 0}{\rm tr}\left(\Pi^f_{\lambda}(\bm 0,\bm x')\right).
\end{align}
The derivation of the spectral density for spinors is more involved, and we just use the result of \cite{Camporesi:1995fb},
\begin{align}
    \rho^f_{H^3}(\lambda)=\frac{\lambda^2+\frac14}{\pi^2 r_K^3}.
\end{align}
This formula will be used when we compute the one-loop corrections and the energy of spinor fields.

We now return to the 5D action~\eqref{2spinordecomp}. We now expand the spinor fields as
\begin{align}
    \hat{\psi}_n(x)=&\int_0^\infty d\lambda\sum_{S,j,m}\left[\hat{b}^S_{n\lambda jm}f^S_{n,\lambda}(t)\Theta^S_{\lambda,j,m}(\chi,\theta,\varphi)-\hat{d}^{S\dagger}_{n\lambda j m}\left(g^S_{n,\lambda}(t)\right)^*\Theta^S_{\lambda,j,m}(\chi,\theta,\varphi)\right],\\
    \hat{\zeta}_n^\dagger(x)=&\int_0^\infty d\lambda\sum_{S,j,m}\left[\hat{b}^S_{n\lambda jm}g^S_{n,\lambda}(t)\Theta^S_{\lambda,j,m}(\chi,\theta,\varphi)+\hat{d}^{S\dagger}_{n\lambda j m}\left(f^S_{n,\lambda}(t)\right)^*\Theta^S_{\lambda,j,m}(\chi,\theta,\varphi)\right].
\end{align}
The mode functions $f_{n,\lambda}(t)$ and $g_{n,\lambda}(t)$ satisfy
\begin{align}
    \left(\begin{array}{c}\dot{f}^S_{n,\lambda}\\ \dot{g}^S_{n,\lambda}\end{array}\right)=-\ri\left(\begin{array}{cc} \frac{S\lambda}{a r_K}&\bar{M}_n\\ M_n&-\frac{S\lambda}{ar_K}\end{array}\right)\left(\begin{array}{c} f^S_{n,\lambda}\\ g^S_{n,\lambda}\end{array}\right),\label{Fmode}
\end{align}
which follows from the Dirac equation. The canonical anti-commutation relation 
\begin{align}
    \left\{\psi_n(t,\chi,\theta,\varphi),\psi^\dagger_{n'}(t,\chi',\theta',\varphi')\right\}=\frac{\delta_{nn'}\delta(\chi-\chi')\delta(\theta-\theta')\delta(\varphi-\varphi')}{\sqrt{g_3}}
\end{align}
leads to the anti-commutation relation
\begin{align}
    \left\{\hat{b}^S_{n\lambda jm},\hat{b}^{S'\dagger}_{n'\lambda' j'm'}\right\}=\delta_{nn'}\delta_{SS'}\delta(\lambda-\lambda')\delta_{jj'}\delta_{mm'}=\left\{\hat{d}^S_{n\lambda jm},\hat{d}^{S'\dagger}_{n'\lambda' j'm'}\right\},
\end{align}
assuming the normalization condition
\begin{align}
    \left|f_{n,\lambda}^S(t)\right|^2+\left|g_{n,\lambda}^S(t)\right|^2=1,
\end{align}
which holds at any $t$ once imposed as an initial condition for the solution to \eqref{Fmode}.

\section{Thermal Contributions to the Energy-Momentum Tensor}\label{appB}
Here we compute the thermal contributions to the energy-momentum tensor in~\eqref{rhophi},\eqref{rhopsi},\eqref{Jphi},\eqref{Jpsi}. First, we consider the scalar energy density
\begin{align}
    \langle\rho_\phi\rangle_{\beta}|_{\rm th}=&{\cal F}_1+{\cal F}_2,\\
    {\cal F}_1=&\frac{1}{a^3(t)(2\pi R)}\sum_i\sum_n\int_0^\infty\frac{dq q^2}{2\pi^2 r_K^3}n_B(\omega_{(i)nq}(t_{\rm ini}))\omega_{(i)nq}(t),\\
    {\cal F}_2=&\frac{3\dot{H}}{4a^3(t)(2\pi R)}\sum_i\sum_n\int_0^\infty\frac{dq q^2}{2\pi^2 r_K^3}\frac{n_B(\omega_{(i)nq}(t_{\rm ini}))}{\omega_{(i)nq}(t)}.
\end{align}
In the following, we omit the sum over species $i$ for notational simplicity. We show several mathematical formulas,
\begin{align}
    e^{-A\sqrt{X}}=&\frac{A}{2\sqrt\pi}\int^\infty_0 duu^{-\frac32}\exp\left[-uX-\frac{A^2}{4u}\right],\label{formula1}\\
    \sqrt{Y}=&-\frac{1}{2\sqrt{\pi}}\int_0^\infty ds s^{-\frac32}(e^{-sY}-1),\label{formula2}
\end{align}
for $A,X,Y>0$. Also, the Bose-Einstein distribution is written as $n_B(\omega)=\sum_{l=1}^\infty e^{-l\beta\omega}$. Using these formulas, we find
\begin{align}
    {\cal F}_1=&-\frac{\tilde\beta}{32\pi^{\frac72}R^5b^{\frac32}a^3}\int_0^\infty du\int_0^{\infty}ds u^{-\frac32}s^{-\frac32}\sum_{l=1}^\infty\left(le^{-\frac{\tilde\beta^2l^2}{u}}\right)\nonumber\\
    &\times\left[(u+\gamma s)^{-\frac32}e^{-\mu_0^2u-\mu^2 s}\vartheta(u+s)-u^{-\frac32}e^{-\mu_0^2u}\vartheta(u)\right]
\end{align}
where we have assumed that $b(t_{\rm ini})=a(t_{\rm ini})=1$ and introduced $\gamma=\frac{b^3}{a^2}$, $\tilde\beta=\frac{\beta}{2R}$,
\begin{align}
    \mu_0^2=& R^2\left( m^2+\frac{1}{r_K^2}-\frac94H^2(t_{\rm ini})-\frac32\dot{H}(t_{\rm ini})\right),\\
    \mu^2=&R^2b^3\left(\frac{m^2}{b(t)}+\frac{1}{a^2(t)r_K^2}-\frac94H^2(t)-\frac32\dot{H}(t)\right),\\
    \vartheta(x)=&\sum_{n\in \mathbb Z}e^{-xn^2},
\end{align}
Note that we have rescaled the integration variables in deriving the above expression. It is more convenient to use $r=s/u$, which leads to
\begin{align}
    {\cal F}_1=&-\frac{\tilde\beta}{32\pi^{\frac72}R^5b^{\frac32}a^3}\int_0^\infty du u^{-\frac72}e^{-\mu_0^2u}\sum_{l=1}^\infty\left(le^{-\frac{\tilde\beta^2l^2}{u}}\right)\int_0^{\infty}drr^{-\frac32}\left[(1+\gamma r)^{-\frac32}e^{-\mu^2 ru}\vartheta((1+r)u)-\vartheta(u)\right]\nonumber\\
   =& -\frac{\tilde\beta^{-4}}{32\pi^{\frac72}R^5b^{\frac32}a^3}\int_0^\infty du u^{-\frac72}e^{-\mu_0^2\tilde\beta^2u}\sum_{l=1}^\infty\left(le^{-\frac{l^2}{u}}\right)\nonumber\\
   &\hspace{2cm}\times\int_0^{\infty}drr^{-\frac32}\left[(1+\gamma r)^{-\frac32}e^{-\mu^2 \tilde{\beta}^2ru}\vartheta((1+r)\tilde\beta^2u)-\vartheta(\tilde\beta^2u)\right].\label{F1integral}
\end{align}

We apply various approximations in the following. First, we divide the $u$-integral region as $[0,1]$ and $[1,\infty]$ and approximate the thermal sum function as
\begin{align}
    \sum_{l=1}^\infty le^{-\frac{l^2}{u}}\approx\left\{\begin{array}{c} e^{-\frac1u}\qquad \text{for }u<1\\ \frac{u}{2}\qquad \text{for }u>1\end{array}\right.
\end{align}
which follows from the series form or asymptotic expansion for $u\gg1$. Although this is a rough approximation, it is enough to obtain the leading order terms. We also need to simplify the KK sum function $\vartheta(x)$, and we note that the Poisson resummation yields $\vartheta(x)=\sum_{w\in \mathbb Z} \sqrt{\frac{\pi}{x}}e^{-\pi^2 w^2/x}$, and therefore we approximate
\begin{align}
    \vartheta(x)\approx \left\{\begin{array}{c} 1 \qquad \text{for } x>1\\ \sqrt{\frac\pi x} \qquad \text{for } x<1\end{array}\right.
\end{align}
where we have taken only the lowest order. Accordingly, we divide the integral regions into several pieces and evaluate the integrals. We consider the parameter regime where the initial temperature is higher than KK mass scale, which implies $\tilde\beta <1$ while the bulk mass or curvature scale is smaller than KK mass scale $\mu<\mu_0<1$. 

We sketch how we approximately evaluate the integral~\eqref{F1integral}. First, we divide the $u$-integration into $[0,1]$, $[1,\tilde\beta^{-2}/2]$, $[\tilde\beta^{-2}/2,\tilde{\beta}^{-2}]$, $[\tilde\beta^{-2},\infty]$. Only within the first region, $\sum_{l=1}^\infty le^{-l^2/u}\approx e^{-1/u}$ and we approximate $\sum_{l=1}^\infty le^{-l^2/u}\approx u/2$ for other regions. We also divide the $r$-integration region to approximate $\vartheta(x)$. For $u\in[0,\tilde\beta/2]$, we define $r_c(u)=\tilde\beta^{-2}u^{-1}-1 (>1)$ and divide the $r$-integration range into $[0,1]$, $[1,r_c]$, $[r_c,\infty]$. For $r\in[0,1]$, we Taylor-expand the integrand in the square bracket with respect to $r$ and take the lowest order in $r$. For $u\in[\tilde\beta^{-2}/2,\infty]$ we divide $r$-integration range into $[0,1]$, $[1,\infty]$ and apply the same approximation for the first region. One would notice that for the high temperature expansion with respect to $\tilde\beta<1$, the leading contribution comes from small $u$ region. We also note that in deriving the leading order expression, we approximate $(1+r)^n\sim r^n$ for $r>1$ while keeping $(1+\gamma r)^n$ since $\gamma$ is time dependent and may become small. 

On the basis of the above approximation, we perform the integration as follows. Here we show only the leading order coming from $u\in [0,1], \ [1,\tilde\beta^{-2}/2]$. The first integral is approximated as
\begin{align}
    &\int_0^1 u^{-\frac72} e^{-\mu_0^2\tilde\beta^2u} e^{-\frac1u}
    \Biggl\{\int_0^1dr r^{-\frac12}\left[\tilde\beta^2 u\theta'\left(\tilde\beta^2u\right)-\left(\frac32\gamma+\mu^2\tilde\beta^2 u\right)\theta\left(\tilde\beta^2u\right)\right]\nonumber\\
    &+\sqrt{\pi}u^{-\frac12}\int_1^{r_c}dr r^{-\frac32}\left[(1+\gamma r)^{-\frac32}(1+r)^{-\frac12}e^{-\mu^2\tilde{\beta}^2ru}\right]\nonumber\\
    &+\int_{r_c}^\infty dr r^{-\frac32}(1+\gamma r)^{-\frac32}e^{-\mu^2\tilde\beta^2ru}-\sqrt{\pi}u^{-\frac12}\int_1^\infty dr r^{-\frac32}\Biggr\}\ \label{Integral}
\end{align}
where we have divided the $r$-integral and used the approximations mentioned above. Now, we approximate the first line as
\begin{align}
    &\int_0^1 u^{-\frac72} e^{-\mu_0^2\tilde\beta^2u} e^{-\frac1u}\int_0^1dr r^{-\frac12}\left[\tilde\beta^2 u\theta'\left(\tilde\beta^2u\right)-\left(\frac32\gamma+\mu^2\tilde\beta^2u\right)\theta\left(\tilde\beta^2u\right)\right]\nonumber\\
    \approx&-\sqrt\pi\tilde\beta^{-1}\left[(1+3\gamma)\int_0^1 u^{-4} e^{-\mu_0^2\tilde\beta^2u} e^{-\frac{1}{u}}+2\mu^2\tilde\beta^2\int_0^1 u^{-3} e^{-\mu_0^2\tilde\beta^2u} e^{-\frac{1}{u}}\right]\nonumber\\
    \approx& -\frac{5\sqrt\pi}{e}\tilde\beta^{-1}\left(1+3\gamma\right)-\frac{4\sqrt\pi}{e}\mu^2\tilde\beta
\end{align}
where we have used $e^{-\mu_0^2\tilde\beta^2u}\approx 1$ in the last line, which can be justified if $\mu_0^2\tilde\beta^2<1$ as the integration range is $u\in[0,1]$. Next, we consider the second line of \eqref{Integral},
\begin{align}
    &\sqrt{\pi}\tilde\beta^{-1}\int_0^1 u^{-4} e^{-\mu_0^2\tilde\beta^2u} e^{-\frac1u}\int_1^{r_c}dr r^{-\frac32}\left[(1+\gamma r)^{-\frac32}(1+r)^{-\frac12}e^{-\mu^2\tilde{\beta}^2ru}\right]\nonumber\\
    \approx&\sqrt{\pi}\tilde\beta^{-1}\int_0^1 u^{-4} e^{-\frac1u}\int_1^{r_c}dr r^{-2}\left[(1+\gamma r)^{-\frac32}\right]\nonumber\\
    \approx&\sqrt{\pi}\tilde\beta^{-1}\int_0^1 u^{-4} e^{-\frac1u}\int_1^{\infty}dr r^{-2}\left[(1+\gamma r)^{-\frac32}\right]\nonumber\\
    =&\frac{5\sqrt\pi}{e}\tilde\beta^{-1}f(\gamma)
\end{align}
where in the second equality we have approximated $(1+r)^{-\frac12}\approx r^{-\frac12}$ and $e^{-\mu^2\tilde{\beta}^2ru}\approx 1$ which can be justified as long as $\mu^2<1$ since the maximal of the exponent is $\tilde\beta^2\mu^2ur_c=\mu^2(1-u\tilde\beta^2)\approx \mu^2$, and in the third equality, we have extended integral region since the integral is localized for smaller $r$. We have also defined the function $f(\gamma)=\frac{1+3\gamma}{\sqrt{1+\gamma}}-3\gamma\log\left(\sqrt{1+\frac{1}{\gamma}}+\frac{1}{\sqrt\gamma}\right)$, which monotonically decreases from $f(0)=1$ for larger $\gamma$. For the third line of \eqref{Integral}, we find
\begin{align}
   & \int_0^1 u^{-\frac72} e^{-\mu_0^2\tilde\beta^2u} e^{-\frac1u}\left\{\int_{r_c}^\infty dr r^{-\frac32}(1+\gamma r)^{-\frac32}e^{-\mu^2\tilde\beta^2ru}-\sqrt{\pi}u^{-\frac12}\int_{1}^\infty dr r^{-\frac32}\right\}\nonumber\\
\approx &-\frac{10\sqrt\pi}{e} \tilde\beta^{-1} 
\end{align}
where the first term in the first line of the curly bracket is neglected since the integrand and contribution from the $r$-integral region under consideration should be smaller than the contribution neglected in the previous integral. Combining these expressions, we obtain
\begin{align}
    \text{\eqref{Integral}}\approx \frac{15\sqrt\pi}{e}\tilde\beta^{-1}\left(1+\gamma-\frac13f(\gamma)\right)+\frac{4}{e}\mu^2\tilde\beta
\end{align}
We apply the same approximation to $u\in [1,\tilde\beta^{-2}/2]$ and the $r$-integration part is computed in the same manner but the $u$-integrand becomes changed due to the thermal sum part, which would give a different overall factor. In this way, we have obtained the following approximate expression 
\begin{align}
    {\cal F}_1\approx\frac{\tilde\beta^{-5}}{32\pi^{3}R^5b^{\frac32}a^3}\left( c_1\left(1+\gamma-\frac13f(\gamma)\right)+c_2\tilde\beta^2\mu^2\right)\label{F1approx}
\end{align}
where we have introduced numerical constants $c_1=\frac{15}{e}+\frac{3}{4}$ and $c_2=\frac{4}{e}+1$. Here we display the leading contributions associated with distinct
time-dependent structures. Therefore, Eq.~\eqref{F1approx} should not be understood as a uniform expansion of $F_1$ in $\tilde\beta$. In the main text, we ignore the $\gamma$-correction terms as they are supposed to be small. We have omitted the terms proportional to $\mu_0$, since $\mu_0$ is time-independent and such terms do not introduce additional time dependence. Note that in deriving the above expression, we have applied multiple approximations, but it is summarized as $$ \mu_0^2\tilde\beta^2<1, \ \mu^2<1.$$ Note also that even if the second condition is violated, it would just mean that the term with $f(\gamma)$ is further suppressed. In the main text, we neglect the term since $f(\gamma)$ soon approaches to zero. Therefore, the violation of the condition does not affect the dynamics much.

We should also emphasize that the numerical coefficients and the detailed shape of $f(\gamma)$ depend sensitively on the sequence of approximations used above, and therefore should not be assigned particular physical
significance. The term linear in $\gamma \equiv b^3/a^2$ arises from a Taylor expansion of the integrand in the small-$r$ region. This expansion is reliable only when $\gamma r \ll 1$, and hence is not uniform in $\gamma$. Therefore, while the linear $\gamma$-dependence may be interpreted as a plausible small-$\gamma$ correction, Eq.~\eqref{F1approx} should not be used to infer the true large-$\gamma$ asymptotic behavior. One may be able to improve the analytical approximation, but we do not pursue this further here, since
it is not the main focus of this work.

Let us compute ${\cal F}_2$. Here we use the following trick. We formally replace $m^2_{i}/b$ as a parameter $s$ in $\omega_{(i)nq}$. Then, noting that $\partial_s\omega_{(i)nq}(t)=\frac{1}{2\omega_{(i)nq}}$. Therefore, by replacing $m^2/b$ in $\mu^2$ by $s$ and differentiating ${\cal F}_1$ with respect to $s$, we find
\begin{align}
    {\cal F}_2=\frac{3\dot H}{2}\partial_s{\cal F}_1|_{s=m^2/b}\approx\frac{3c_2\dot Hb^{\frac32}\tilde\beta^{-3}}{64\pi^{3}R^3a^3}.
\end{align}
A similar trick applies to the source term $\langle\hat J_{\phi}\rangle$~\eqref{Jphi}. Noting that $\partial_b\omega_{(i)nq}=-\frac{3}{2b\omega_{(i)nq}}\left(\frac{n^2}{R^2b^3}+\frac{m^2_i}{3b}\right)$, we find
\begin{align}
    \langle\hat{J}_{\phi}\rangle_\beta|_{\rm th}=-\frac{2b^2}{3}\partial_b{\cal F}_1,
\end{align}
which implies that the source term can be derived as the variation of the thermal effective potential ${\cal F}_1$ with respect to the radion as expected.

We proceed to the spinor field case. The thermal contribution of spinor fields to energy density is given by
\begin{align}
  {\cal F}_3=\frac{2}{a^3(t)(2\pi R)}\sum_{I}\sum_{n\in\mathbb Z}\int_0^\infty d\lambda\frac{\lambda^2+\frac14}{\pi^2r_K^3}\omega_{(I)n\lambda}(t)n_{F}(\omega_{(I)n\lambda}(t_{\rm ini})).
\end{align}
We omit the sum and indices $(I)$ in the following. With the integral representation formulas~\eqref{formula1}, \eqref{formula2}, we can write ${\cal F}_3={\cal F}_{31}+{\cal F}_{32}$ as
\begin{align}
    {\cal F}_{31}=&-\frac{\tilde\beta^{-4}}{16\pi^{\frac72}R^6a^3b^{\frac32}}\int_0^\infty duu^{-\frac72}e^{-\nu_0^2\tilde\beta^2u}\sum_{l=1}^{\infty}(-1)^{l+1}e^{-\frac{l^2}{u}}\nonumber\\
    &\qquad \times\int_0^\infty dr r^{-\frac32}\left[(1+\gamma r)^{-\frac32}e^{-\nu^2\tilde\beta^2 r u}\Theta((1+r)\tilde\beta^2u)-\Theta(\tilde\beta^2u)\right],\\
    {\cal F}_{32}=&-\frac{\tilde\beta^{-2}}{32\pi^{\frac72}R^4a^3b^{\frac32}r_K^2}\int_0^\infty duu^{-\frac52}e^{-\nu_0^2\tilde\beta^2u}\sum_{l=1}^{\infty}(-1)^{l+1}e^{-\frac{l^2}{u}}\nonumber\\
    &\qquad \times\int_0^\infty dr r^{-\frac32}\left[(1+\gamma r)^{-\frac12}e^{-\nu^2\tilde\beta^2 ru}\Theta((1+r)\tilde\beta^2u)-\Theta(\tilde\beta^2u)\right],
\end{align}
where $\nu_0^2=m^2R^2$, $\nu^2=m^2R^2b^2$. The integral ${\cal F}_{31}$ is almost same as the scalar case ${\cal F}_1$ except for the overall factor $4$ originating from the number of degrees of freedom in a 5D Dirac field, and the factor $(-1)^{l+1}$ associated with the statistical property. 

We apply the same approximation to evaluate the integral analytically. Note that the thermal sum function has different asymptotic behavior as
\begin{align}
    \sum_{l=1}^{\infty}(-1)^{l+1}e^{-\frac{l^2}{u}}=\left\{\begin{array}{c} e^{-\frac1u} \qquad \text{for }u\ll1\\ \frac12 \qquad \text{for }u\gg1\end{array}\right.
\end{align}
where we have taken the leading order in the asymptotic expansion. In the high-temperature expansion with respect to $\tilde\beta\ll1$, the leading order contribution appears from $u\ll1$ where the thermal sum function behaves in the same way as the scalar field case. This is reasonable as the statistical difference appears in the limit. 
Therefore, we obtain the leading order expression for ${\cal F}_{31}$ to be 
\begin{align}
{\cal F}_{31}\approx \frac{\tilde\beta^{-5}}{8\pi^{3}R^5b^{\frac32}a^3}\left( c_1\left(1+\gamma-\frac13f(\gamma)\right)+c_2\tilde\beta^2\nu^2\right).
\end{align}
The difference from \eqref{F1approx} is the last term and the overall factor. For ${\cal F}_{32}$, we apply the same approximations, which yields
\begin{align}
    {\cal F}_{32}=\frac{\tilde\beta^{-3}}{32\pi^{3}R^3a^3b^{\frac32}r_K^2}\left(c_3\left(3+\frac12\gamma-g(\gamma)\right)+c_4\tilde\beta^2\nu^2\right),
\end{align}
where $c_3=\frac2e+\frac14$, $c_4=\frac5e+\frac16$ and $g(\gamma)=\sqrt{1+\gamma}-\gamma\log\left(\sqrt{1+\frac1\gamma}+\frac{1}{\sqrt\gamma}\right)$. For the spinor source term, $\langle \hat{J}_\Psi\rangle$, by the same reason as the scalar case, we can find
\begin{align}
    \langle\hat{J}_{\Psi}\rangle_\beta|_{\rm th}=-\frac{2b^2}{3}\partial_b{\cal F}_3.
\end{align}
Thus, we have derived the thermal contributions to the energy density in the high temperature limit. We note that the thermal energy in higher-dimensional models and its behavior are studied e.g. in \cite{Yoshimura:1984qa,Maeda:1985xba,Otsuka:2022vgf,Otsuka:2024xsp}. Our result is different from \cite{Yoshimura:1984qa,Maeda:1985xba} because they considered the case where the scale factors are periodic in the imaginary time such that they also maintain the thermal equilibrium approximation. Regarding~\cite{Otsuka:2022vgf,Otsuka:2024xsp}, the authors considered $S^2$ compactification and the structure of the KK masses is different from ours.

We do not attempt to make a precision computation of the thermal effective potential in this toy model. 
The thermal terms are used to capture the leading qualitative effect of an initially excited KK population on the radion dynamics. 
The important point for our purpose is that these contributions are diluted before the radion reaches the late-time compactified minimum, while the curvature component can still provide sufficient Hubble friction. 
Subleading corrections to the thermal approximation may shift the detailed numerical trajectory, but within the qualitative toy-model regime considered here, they are not expected to alter the basic curvature-assisted trapping mechanism discussed in the main text.

\section{Cutoff Scale}\label{appC}
In our discussion, multiple scales and corresponding regimes appear. We therefore describe the meaning of each scale and how the effective description of each regime may break down. We follow the discussion e.g. in \cite{Han:1998sg,Cheng:2010pt}.

First, we have 5D Planck scale $M_5$ but within our coordinate system, the local proper time is given by $d\tau=\frac{1}{\sqrt{b}}dt$, and then, the proper frequency $e^{-\ri E\tau}=e^{-\ri (E/\sqrt{b}) t}$. For the classical description of 5D gravity, we need to impose $M_5>E$, but this implies that the energy cutoff with our time $t$ is $E_{\rm cut}(t)=M_5/\sqrt{b(t)}$. Correspondingly, the initial temperature $T$ should be
\begin{align}
    T\leq \frac{M_5}{\sqrt{b(t_{\rm ini})}}.
\end{align}
Above the energy scale $E_{\rm cut}(t)=M_5/\sqrt{b(t)}$, quantum gravitational effects become important and we cannot trust the classical background approximation. 

We also have 4D Planck scale $M_{\rm pl}^2=2\pi R M_5^3$. Recall that, the parameter $R$ is a reference value of the compact space radius and not a physical parameter. Also, $M_{\rm pl}$ is greater than $M_5$ if we take roughly $R>M_{5}^{-1}$. In the naive 4D EFT, the 4D Planck scale appears to set the quantum-gravity scale, which can be much greater than the 5D gravitational scale $M_5/\sqrt{b(t)}$. Then, which is the appropriate quantum gravity scale?

The key to resolving the above question is to notice that 4D EFT description breaks down above the KK scale $m_{\rm KK}(t)=\frac{1}{Rb^{\frac32}(t)}$. Let us estimate at which scale the strong coupling appears. Given the center-of-mass energy $E$, a generic $2\to2$ scattering amplitude mediated by graviton interactions is given by
\begin{align}
    {\cal A}\propto \left(\frac{E^2}{M_{\rm pl}^2}\right).
\end{align}
However, we need to take into account the number of KK modes contributing to the scattering. The number of KK modes below the energy $E$ is estimated as
\begin{align}
    N(E)=\frac{E}{m_{\rm KK}(t)}=ERb^{\frac32}(t).
\end{align}
Therefore, the total amplitude taking into account the number of KK modes is
\begin{align}
    {\cal A}_{\rm tot}\propto {\cal A}N(E)\propto \left(\frac{E^2}{M_{\rm pl}^2}\right)\times ERb^{\frac32}=\frac{E^3}{2\pi (M_5/\sqrt b)^3}.
\end{align}
Thus, this scaling shows that the actual quantum gravity scale within our setup is $M_5/\sqrt{b}$ rather than 4D Planck scale,
 because 4D EFT already breaks down at $m_{\rm KK}=1/(Rb^{\frac32})$.

We require that the initial KK scale should be smaller than the cutoff scale, which implies, taking $b(t_{\rm ini})=1$,
\begin{align}
    M_5>\frac{1}{R}\Leftrightarrow M_5R>1.
\end{align}
Then, the 4D Planck scale must be larger than the 5D Planck scale
\begin{align}
    M_{\rm pl}^2=2\pi RM_5^3>M_5^2.
\end{align}
Then, within our scenario, we consider the parameter regime schematically described as
\begin{align}
    M_{\rm pl}>M_5>T>\frac{1}{R}>m_{\rm rad}>H_{\rm inf}
\end{align}
where $T$ is the initial thermal state temperature, $m_{\rm rad}$ the (canonical) radion mass at the local minimum, and $H_{\rm inf}$ the 4D inflation Hubble scale. We also take the initial scale factor $a(t_{\rm ini})=1$, and then the curvature energy scale should also be below the cutoff scale, which implies
\begin{align}
    M_5^4>\frac{3M_{\rm pl}^2}{r_K^2}.
\end{align}

\section{Effective Frequency of a Scalar Field in the Open FRW Cosmology}\label{appD}
The effective frequency of scalar fields within our background is given by
\begin{align}
    \omega_{nq}^2=\frac{q^2+1}{a^2r_K^2}+\frac{m^2}{b}+\frac{n^2}{R^2b^3}-\frac{9}{4}H^2-\frac32\dot{H}.
\end{align}
The last two terms may in principle make the effective frequency squared negative. 
Let us consider a background filled with a fluid with energy density $\rho$ and pressure $p=w\rho$. 
The Einstein equations read
\begin{align}
H^2 &= \frac{\rho}{3M_{\rm pl}^2} + \frac{1}{a^2 r_K^2}, \\
\dot H &= -\frac{\rho+p}{2M_{\rm pl}^2} - \frac{1}{a^2 r_K^2}.
\end{align}
Then, the combination appearing in the effective frequency is
\begin{align}
\frac{1}{a^2 r_K^2} - \frac{9}{4}H^2 - \frac{3}{2}\dot H
= \frac{3w}{4M_{\rm pl}^2}\rho + \frac{1}{4a^2 r_K^2}.
\end{align}
Therefore, for $w>0$, this contribution is manifestly positive. 
In particular, in our setup, once the positive bulk mass and KK mass terms are included, the effective frequency can remain non-tachyonic. 
We also note that even if $\omega_{nq}^2$ becomes negative for sufficiently long-wavelength modes, this does not necessarily imply a genuine tachyonic instability, but may simply indicate that the corresponding mode is superhorizon.
\bibliographystyle{JHEP}
\bibliography{main.bib}
\end{document}